\documentclass[11pt]{article}
\pdfoutput=1

\usepackage{jheppub}

\usepackage{supertabular} 
\usepackage{indentfirst} 
\usepackage{oldgerm}
\usepackage{textfit}
\usepackage{eso-pic,graphicx}
\usepackage[amssymb]{SIunits} 
\usepackage{nicefrac} 
\usepackage{footnote}
\usepackage{latexsym,amsfonts,mathrsfs,bbold,mathtools,esint}
\usepackage{nicefrac} 
\usepackage{booktabs} 
\usepackage{array} 
\usepackage{paralist} 

\newcommand{\diag}{\mathrm{diag}\:}

\newcommand{\csch}{\mathrm{csch}\:}

\DeclareMathOperator{\si}{si}

\newcommand{\delslash}{{\partial\mkern-9mu/}} 
\newcommand{\dslash}[1]{{#1}\mkern-9mu/} 

\begin{document}

\title{\bf Quantum integrability of the Alday-Arutyunov-Frolov model} 
\author[a,b]{A. Melikyan}
\author[a,b]{A. Pinzul}
\author[c]{V. O. Rivelles}
\author[c]{G. Weber}
\affiliation[a]{ \sl{Instituto de F\'{\i}sica},
\sl{Universidade de Bras\'{\i}lia, \\70910-900, Bras\'{\i}lia, DF, Brasil}}
\affiliation[b]{\sl{{International Center of Condensed Matter Physics}},\\
\sl{C.Postal 04667, Brasilia, DF, Brazil}}
\affiliation[c]{\sl{Instituto de F\'{\i}sica},
\sl{Universidade de S\~{a}o Paulo,\\ C. Postal 66318, 05315-970, S\~{a}o Paulo, SP, Brasil}}

\emailAdd{amelik@gmail.com}
\emailAdd{apinzul@unb.br}
\emailAdd{rivelles@fma.if.usp.br}
\emailAdd{weber@fma.if.usp.br}

\date{}

\abstract{We investigate the quantum integrability of the Alday-Arutyunov-Frolov (AAF) model by calculating the three-particle scattering amplitude at the first non-trivial order and showing that the S-matrix is factorizable at this order. We consider a more general fermionic model and find a necessary constraint to ensure its integrability at quantum level. We then show that the quantum integrability of the AAF model follows from this constraint. In the process, we also correct some missed points in earlier works. }

\arxivnumber{1106.0512}
\maketitle


\section{Introduction}

The AdS/CFT correspondence \cite{Maldacena:1997re} has provided deep insight into the intricate dynamics of both gauge and string theories, much of which can be ascribed to the uncovering of integrable structures on both sides of the duality \cite{Minahan:2002ve,Beisert:2003jj,Kazakov:2004qf,Gromov:2006cq,Arutyunov:2004vx,Arutyunov:2003rg}\footnote{For a comprehensive review of the role of integrability in the context of the AdS/CFT correspondence, see  \cite{Beisert:2010jr}.}. On the string theory side, the classical integrability of the sigma-model describing the superstring on $AdS_5\times S^5$ is relatively well understood \cite{Bena:2003wd}. Though, in order to better understand the AdS/CFT conjecture it is necessary to quantize the superstring theory. This, however, has not yet been achieved using the conventional methods developed in the context of quantum integrable systems.

It is, nevertheless, feasible and interesting to study string theory truncated to smaller subsectors which are dual to closed sectors of the gauge theory \cite{Beisert:2004ry,Staudacher:2004tk,Alday:2005gi,Arutyunov:2004yx,Alday:2005jm}. Even though such reduced models may lose some important properties of the full theory, such as conformal invariance, they are still expected to be classically integrable, providing simpler but nonetheless representative examples of the difficulties associated with the quantization of the superstring theory on $AdS_5 \times S^5$. One important case is the Alday-Arutyunov-Frolov (AAF) model \cite{Alday:2005jm}. It arises in this context as the consistent truncation of the superstring theory in the uniform gauge to the $\mathfrak{su}(1|1)$ sector. It is interesting also to note that the AAF model also appears in a subsector of type IIA superstrings on $AdS_4 \times \mathbb{CP}^3$ \cite{Dukalski:2009pr}.

The AAF model is a particularly interesting example of a classically integrable model in several aspects. It is the first non-trivial purely fermionic integrable model which is highly non-linear and singular\footnote{By singularity we mean the presence of derivatives in the interaction Hamiltonian which results in a very singular quantum mechanical Hamiltonian and associated quantum conserved charges.} when compared to the standard case of the fermionic Thirring model. Such singularities in the context of integrable systems are hard to handle. One example where such difficulties are present is the Landau-Lifshitz model \cite{Melikyan:2008ab,Melikyan:2010bi,Melikyan:2010fr}, which is generated by a string in the $\mathfrak{su}(2)$ subsector. It has been shown that the complete understanding of the quantum inverse scattering method is only possible upon careful analysis of the singularities, the associated quantum operators, and the reconstruction of the correct Hilbert space. Another intriguing aspect of the AAF model is its non-linear structure. Unlike the simpler Thirring model where the interaction vertex is of fourth order in the fermionic fields, the AAF model contains also a sixth order interaction term. As we will see, this introduces further obstacles in the analysis of quantum integrability. It is the purpose of this paper to understand the interplay among the several types of interactions which in the end results in the quantum integrability of the model. 

The AAF model inherits the classical integrability of the superstring theory on $AdS_5 \times S^5$, as the Lax representation of the full string sigma-model admits the same consistent truncation \cite{Alday:2005jm}. This classical integrability was then conjectured to hold at quantum level by \cite{Klose:2006dd}, where the corresponding Bethe equations were derived from the knowledge of the two-particle $S$-matrix and the assumption of the S-matrix factorization. In this case the AAF model was regarded as a two-dimensional field theory and its $S$-matrix was computed by perturbative methods. 

The quantum inverse scattering method is the only reliable and desirable method to account for all non-perturbative effects in an integrable model. Unfortunately, it has not yet been developed for the AAF model due to its singular and non-linear nature. In this situation, the perturbative approach is essentially the only available method to probe quantum integrability and to obtain the Bethe equations without serious technical problems. It is clear, however, that within the perturbative quantum field theoretic approach non-perturbative information may be lost. Still, this does not happen in many known integrable models, for which the quantum inverse scattering method leads to the same results as the perturbative calculations. This happens  because, in such cases, the S-matrix can be found exactly to all orders.  We mention here another interesting aspect of the AAF model which sets it apart from all other known classical integrable models. Namely, the highly non-linear nature of its Poisson structure, which extends up to the sixth order in the fermions and its spatial derivatives. In perturbation theory this information is not essential, but it is very desirable to understand its effect within the quantum inverse scattering formalism. Thus, until the quantum inverse scattering is fully developed and understood, perturbative calculations are essentially the only working tool at our disposal.

There is, however, an alternative formulation of the AAF model, as explained in the original work \cite{Alday:2005jm}, since it is possible to perform a field redefinition to trivialize the Poisson structure at the price of getting a complicated Hamiltonian. Nonetheless, both approaches are plagued with the usual problems found in the quantization process of continuous integrable models, which reflect the ill-defined operator product at the same point. Although there exists standard discretization techniques that can, in principle, be used to avoid this problem, they usually lead to very complicated results and, more importantly, they might not be readily applicable to the more involved string model on $AdS_5 \times S^5$.  Thus, it is desirable to deal directly with the continuous AAF model, which should shed some light on possible ways to overcome the fundamental quantization difficulties of the full string model.

Yet another possibility is to find an alternative gauge choice that linearizes the equations of motion. Actually, if the string model truncated to the $\mathfrak{su}(1|1)$ sector has its reparametrization invariance fixed by means of the uniform light-cone gauge \cite{Arutyunov:2005hd}, it becomes a two-dimensional theory for free massive Dirac fermions. In this case the quantization is trivial and the spectrum can be easily obtained. However, it is not clear whether the classical equivalence between the AAF model in the uniform gauge and  in the uniform light-cone gauge survives quantization. The reason for that lies in the fact that the conformal invariance, which is necessary for quantum gauge equivalence, is broken in the reduction to the classically closed sector. That being the case, it is an interesting question to quantize the AAF model in the uniform gauge and compare its spectrum with the one obtained from the free action in \cite{Arutyunov:2005hd}.

In this paper we  probe the quantum integrability of the AAF model in the uniform gauge by analyzing the factorizability of its $S$-matrix. We proceed along the lines of our earlier work \cite{Melikyan:2008cy} and consider the three-particle scattering within the framework of quantum field theory. It is important to bear in mind that a necessary condition for a factorizable scattering is the absence of genuine three-particle interactions. Clearly, this is, \emph{a priori}, not the case for the AAF model since it explicitly contains a three-particle interaction vertex. In fact, we show that even in the first non-trivial order, the $S$-matrix factorization property can be verified only if the higher order contributions are taken into account. 

For our approach, it is convenient to assign canonical mass dimensions to the originally dimensionless fields of the AAF model. This naturally leads to introduction of two dimensionful coupling constants, one for each interaction vertex. In this paper we analyze a more general model, treating the coupling constants as independent, and derive a necessary constraint to ensure the \emph{quantum} integrability of the model. This constraint effectively reduces the number of coupling constants to one, which is in complete agreement with the AAF model. Indeed, the original \emph{classical} dimensionless AAF action contains only one parameter $\lambda$, and we show that our general constraint is consistent with this action.

The intricate mechanism behind $S$-matrix factorization is the same unveiled for the Landau-Lifshitz (LL) model in \cite{Melikyan:2008cy}. However, in the AAF case, its verification is not so straightforward due a substantially more complicated diagrammatic analysis. As a result, the daunting perturbative computations make it hard to understand the various cancellations necessary for the quantum integrability of the model. Moreover, some missed factors in the previous literature have been revealed and corrected in this paper. They were found in the process of proving the conditions for quantum integrability. In particular, the Lagrangian for the interacting massive Dirac fermion in two dimensions, originally derived by \cite{Alday:2005jm}, has a missing factor of $\frac{1}{2}$ in front of the three-particle interaction term. Its absence would prevent quantum integrability, as a delicate fine-tuning between the coupling constants of the different interaction vertices is required for $S$-matrix factorization. Even though this missing factor does not affect the two-particle calculations performed in \cite{Klose:2006dd}, there is a further crucial overall sign difference, which leads to the derivation of the inverse $S$-matrix instead of the proper one, changing all the subsequent analysis concerning excited and bound states.

Several other technical subtleties, absent in the much simpler LL case, make the computation of scattering amplitudes a much harder problem for the AAF model. First, the fact that the theory is relativistic invariant demands a quantization with respect to a false vacuum in order to render the propagator purely retarded. Nevertheless, the two poles of the propagator should be carefully taken into account, and are essential in the analysis of the continuity of the scattering amplitudes and in the cancellation of non-integrable contributions. Moreover, the presence of spinorial products requires a great care in the combinatorial analysis, making the higher order calculations quite involved.

The paper is organized as follows. In section {\bf \ref{Overview of AAF}} we give a very brief review of the AAF model incorporating the factor missed by \cite{Alday:2005jm}. In section {\bf \ref{AAF QFT}} we set up the AAF model as a quantum field theory and prepare all the necessary tools for computing the two- and three-particle $S$-matrices. In section {\bf \ref{2 Particle Scattering}} we give another derivation of the two-particle $S$-matrix  based on standard techniques. In section {\bf \ref{3 Particle Scattering}} we present our analysis of the three-particle $S$-matrix and we show its factorization at the first non-trivial order. Finally, we collect some important technical details in the appendices.

\section{Overview of the Alday-Arutyunov-Frolov model} \label{Overview of AAF}

In this section we review the AAF model \cite{Alday:2005jm}, which emerges as a result of the consistent truncation of the superstring sigma model on $AdS_5 \times S^5$ to the $\mathfrak{su}(1|1)$ sector. 
Let us briefly remind the reduction process. 

 The $\mathfrak{su}(1|1)$ sector of superstring theory is defined to be the smallest sector of the full $AdS_5 \times S^5$ theory to contain all the states dual to the operators contained in $\mathfrak{su}(1|1)$ sector of $\mathcal{N}=4$ SYM. The latter consists, in the $\mathcal{N}=1$ language, of gauge invariant composite operators made of products between a complex scalar $Z$ from the scalar supermultiplet and a Weyl fermion $\Psi$ from the gaugino supermultiplet. In order to proceed with the truncation it is necessary to single out a string scalar field to be in correspondence with the field $Z$ from the dual gauge theory, while keeping only the time coordinate from $AdS_5$ non-zero. The residual bosonic symmetry algebra can then be used to decompose the original 16 complex fermions into four sectors, comprising 4 fermions each. It is possible to reduce the superstring equations of motion to one of these sectors, and furthermore set a pair of the remaining fermions consistently to zero. Next, it is tempting to put one of these in direct correspondence with the gauge theory fermion $\Psi$. However, this is not the case, as a consistent truncation which keeps only one fermion non-zero is forbidden by the cubic couplings arising from the Wess-Zumino term in the superstring Lagrangian. It is important to bear in mind that the $\mathfrak{su}(1|1)$ sector of superstring theory does not coincide with the $\mathfrak{su}(1|1)$ closed sector of the dual gauge theory, since, to begin with, they contain a different number of degrees of freedom.

In addition to the conditions discussed above, one must still fix the reparametrization invariance of the superstring action. A fitting way to do this corresponds to imposing the uniform gauge \cite{Arutyunov:2004yx}, which identifies the world-sheet time $\tau$ with the $AdS_5$ global time $t$, and fixes the only non-vanishing component $J=J_3$ of the $S^5$ angular momentum to be equal to the corresponding $\mathfrak{u}(1)$ charge. By imposing the uniform gauge, the two bosonic degrees of freedom remaining after the reduction to the $\mathfrak{su}(1|1)$ sector: the $S^5$ angle $\phi$ corresponding to the scalar $Z$ and the $AdS_5$ global time $t$, are removed. In particular, by solving the constraints introduced by this gauge choice, the angle $\phi$ can now be expressed in terms of the fermionic coordinates.  So that the remaining physical degrees of freedom are purely fermionic. Remarkably, the two complex space-time fermions can be grouped into a single two-component world-sheet Dirac spinor. Accordingly, the action for the truncated model reduces to a non-trivially interacting Lorentz invariant action of the massive Dirac fermion on the flat two-dimensional world-sheet.

The extensive details of the derivation, as well as the  notations which we also follow here, can be found in the original paper \cite{Alday:2005jm}. Our starting point is the classically integrable AAF model Lagrangian (see the expression (5.3) of \cite{Alday:2005jm}): 
\begin{align}
	\label{AAF lagrangian 5.3} \mathscr{L} &= -J - \frac{iJ}{2}\left(\bar{\psi}\rho^0
	\partial_0 \psi - 
	\partial_0 \bar{\psi}\rho^0\psi \right) +i\kappa\left( \bar{\psi} \rho^1
	\partial_1\psi - 
	\partial_1\bar{\psi}\rho^1 \psi \right) + J\bar{\psi}\psi \: + \nonumber \\
	&+ \frac{iJ}{4}\left( \bar{\psi} \rho^0 
	\partial_0 \psi -
	\partial_0 \bar{\psi} \rho^0 \psi\right) \bar{\psi}\psi - \frac{i\kappa}{2} \left( \bar{\psi}\rho^1
	\partial_1\psi - 
	\partial_1 \bar{\psi} \rho^1\psi\right) \bar{\psi}\psi -\frac{J}{2}\left( \bar{\psi} \psi\right)^2 + \nonumber \\
	&+ \frac{\kappa}{2} \epsilon^{\alpha \beta}\left( \bar{\psi} 
	\partial_{\alpha} \psi \; \bar{\psi} \rho^5 
	\partial_{\beta} \psi - 
	\partial_{\alpha} \bar{\psi} \psi \:\; 
	\partial_{\beta} \bar{\psi} \rho^5 \psi \right) + \frac{\kappa}{8} \epsilon^{\alpha \beta} \left(\bar{\psi} \psi \right)^2 
	\partial_{\alpha}\bar{\psi} \rho^5 
	\partial_{\beta} \psi, 
\end{align}
where the Dirac matrices $\rho^0$, $\rho^1$ and $\rho^5$ are defined in appendix {\bf \ref{Appendix Dirac}} in (\ref{dirac matrices rho}), and the Levi-Civita tensor is such that $\epsilon^{01}=\epsilon_{10}=1$. By means of the following field redefinition: 
\begin{equation}
	\psi \to \psi + \frac{1}{4} \psi \left(\bar{\psi}\psi \right), \quad \bar{\psi} \to \bar{\psi} + \frac{1}{4} \bar{\psi} \left(\bar{\psi}\psi \right), 
\end{equation}
one can simplify the Lagrangian (\ref{AAF lagrangian 5.3}) further, and write it in the form: 
\begin{align}
	\label{AAF lagrangian 5.5} \mathscr{L}_{AAF} &= -J - \frac{iJ}{2} \left(\bar{\psi} \rho^0 
	\partial_0 \psi - 
	\partial_0 \bar{\psi} \rho^0 \psi \right) + i \kappa \left(\bar{\psi}\rho^1 
	\partial_1 \psi - 
	\partial_1 \bar{\psi} \rho^1 \psi \right) + J\bar{\psi}\psi + \nonumber \\
	&+ \frac{\kappa}{2} \epsilon^{\alpha \beta} \left( \bar{\psi}
	\partial_{\alpha} \psi \; \bar{\psi} \rho^5 
	\partial_{\beta} \psi -
	\partial_{\alpha}\bar{\psi} \psi \; 
	\partial_{\beta} \bar{\psi} \rho^5 \psi \right) - \frac{\kappa}{8} \epsilon^{\alpha \beta} \left(\bar{\psi}\psi\right)^2 
	\partial_{\alpha}\bar{\psi}\rho^5 
	\partial_{\beta}\psi. 
\end{align}

It is important to emphasize here a significant difference in the last interacting term of our Lagrangian (\ref{AAF lagrangian 5.5}) when compared to the Lagrangian (5.5) of \cite{Alday:2005jm}. The extra factor $\frac{1}{2}$ that appears in our Lagrangian is crucial, as we will show in the subsequent sections, for the quantum integrability of the model, and was missed in \cite{Alday:2005jm}.\footnote{We thank S. Frolov for confirming this correction.} As a result, this missed factor had propagated to \cite{Klose:2006dd}, where the two-particle S-matrix was obtained for the first time. Although essential for the S-matrix factorization of $n \geq 3$ particles scattering process, and consequently for the quantum integrability, this extra factor does not affect the two-particle S-matrix calculation. However, another missed point, as we will explain below, effectively changes the two-particle S-matrix of \cite{Klose:2006dd} to its inverse. 

We stress that all these results have been obtained by analyzing the quantum integrability of the model, which imposes a strong constraint on the coupling constants, and then checking its consistency in the classical limit. This is explained in details in the subsequent sections.

\section{The AAF model as quantum field theory}\label{AAF QFT}

Due to the highly non-trivial form of the Poisson brackets, which extends up to the eighth order in the fermions and their spatial derivatives, it is not an easy task to directly quantize the theory, defined by the Lagrangian (\ref{AAF lagrangian 5.5}), by the standard methods of the quantum inverse scattering method. However, since we are interested in probing the quantum integrability of the AAF model, there is an alternative framework: to consider the AAF model as a quantum field theory and study the factorability of its $S$-matrix. In this section we set up all the necessary tools for computing the two- and three-particle $S$-matrices.

The starting point is the action defined by the Lagrangian (\ref{AAF lagrangian 5.5}):
\begin{equation}
	\label{AAF action 5.5} S = \int d\tau \: \int_0^{2 \pi} \frac{d\sigma}{2\pi} \: \mathscr{L}_{AAF}, 
\end{equation}
which is not, however, explicitly Lorentz invariant. This can be readily fixed if we rescale the world-sheet coordinate $\sigma$: 
\begin{equation}
	\label{sigma shift} \sigma \to - \frac{2 \kappa}{J} \sigma, 
\end{equation}
leading to:
\begin{align}\label{AAF action lorentz invariant} 
	S &= \frac{\kappa}{\pi} \int d\tau\: \int_{-\frac{\pi J}{\kappa}}^0 d\sigma \: \left[ -1 - \frac{i}{2} \left( \bar{\psi} \rho^{\alpha} \partial_{\alpha} \psi - \partial_{\alpha} \bar{\psi} \rho^{\alpha} \psi \right) + \bar{\psi}\psi \:- \right.  \nonumber \\
	&- \left.  \frac{1}{4}\epsilon^{\alpha \beta} \left( \bar{\psi}
	\partial_{\alpha} \psi \; \bar{\psi} \rho^5 \partial_{\beta} \psi - \partial_{\alpha}\bar{\psi} \psi \; \partial_{\beta} \bar{\psi} \rho^5 \psi \right) + \frac{1}{16} \epsilon^{\alpha \beta} \left(\bar{\psi}\psi\right)^2 
	\partial_{\alpha}\bar{\psi}\rho^5 \partial_{\beta}\psi \right]. 
\end{align}
As defined in the Lagrangian (\ref{AAF action lorentz invariant}), the kinetic term has a $(-1)$ sign in front of it, when compared to the standard convention. There are two equivalent ways to develop the perturbation theory. One can work directly with the signs defined in (\ref{AAF action lorentz invariant}), and in this case one has to be careful with the mode expansion (see the next section), since the energy is negative definite now. In other words, the particle and anti-particle operators are switched in comparison with the standard textbook convention. Alternatively, we could modify the Lagrangian and make the kinetic term, together with the energy, positive definite. To achieve this, we can make the transformation $\mathscr{L} \longrightarrow (-1)\mathscr{L}$. On the classical level this transformation does not change the dynamics of the model. The situation on the quantum level is slightly more complicated.  Firstly, the propagator will acquire an additional overall minus sign, and, secondly, the poles will be switched, so that the corresponding positive and negative energy states are interchanged, compared to the first approach. 
Both approaches are equivalent, however, one cannot mix the two, as it appears to be the case in \cite{Klose:2006dd}, and which had led to the inverted $S$-matrix.\footnote{See the discussion at the end of the section {\bf \ref{Sfactor}} on the relation between our results and the results of \cite{Klose:2006dd}.} Here, we choose the second path, and make the kinetic term positive by multiplying the classical Lagrangian by $(-1)$.

Finally, we redefine the integration variable $\sigma \longrightarrow \sigma + \frac{2 \pi J}{\sqrt{\lambda}}$ as in \cite{Klose:2006dd}, neglect the constant term in (\ref{AAF action lorentz invariant}), fix $\kappa = \frac{\sqrt{\lambda}}{2}$, and change the Dirac matrices basis through the similarity transformation (\ref{dirac matrices similarity tranformation}). Then, the action becomes: 
\begin{align}
	\label{AAF S action symmetric} S &= \frac{\sqrt{\lambda}}{2 \pi} \int d\tau \: \int_0^{\frac{2 \pi J}{\sqrt{\lambda}}} d\sigma \: \left[ \frac{i}{2} \left(\bar{\psi} \: \gamma^{\alpha} 
	\partial_{\alpha} \psi - 
	\partial_{\alpha} \bar{\psi} \: \gamma^{\alpha} \psi \right) - \bar{\psi} \psi \: + \right. \nonumber \\
	&+ \left. \frac{1}{4}\epsilon^{\alpha \beta} \left( \bar{\psi}
	\partial_{\alpha} \psi \; \bar{\psi}\: \gamma^3 
	\partial_{\beta} \psi -
	\partial_{\alpha}\bar{\psi} \psi \; 
	\partial_{\beta} \bar{\psi}\: \gamma^3 \psi \right) - \frac{1}{16} \epsilon^{\alpha \beta} \left(\bar{\psi}\psi\right)^2 
	\partial_{\alpha}\bar{\psi}\:\gamma^3
	\partial_{\beta}\psi \right]. 
\end{align}

Up to now, we have been working only with dimensionless quantities. However, for our purposes it is convenient to assign canonical mass dimensions to the fields. To start with, we perform the following coordinate transformation \cite{Klose:2006dd}: 
\begin{equation}
	\label{AAF S assinging mass} x^{\alpha} \to y^{\alpha} = \frac{\sqrt{\lambda}}{2 \pi} x^{\alpha}, \quad \textrm{with} \quad x^0 =\tau, \quad x^1 = \sigma,
\end{equation}
under which the action (\ref{AAF S action symmetric}) becomes: 
\begin{align}
	\label{AAF S mass action} S &= \int dy^0 \: \int _0^J dy^1 \: \left[ i \bar{\psi} \delslash \psi \: - \frac{2\pi}{\sqrt{\lambda}} \bar{\psi} \psi \right. + \nonumber \\
	&+ \left. \frac{\sqrt{\lambda}}{8\pi} \epsilon^{\alpha \beta} \left( \bar{\psi}
	\partial_{\alpha} \psi \; \bar{\psi}\: \gamma^3 
	\partial_{\beta} \psi -
	\partial_{\alpha}\bar{\psi} \psi \; 
	\partial_{\beta} \bar{\psi}\: \gamma^3 \psi \right) - \frac{\sqrt{\lambda}}{32 \pi} \epsilon^{\alpha \beta} \left(\bar{\psi}\psi\right)^2 
	\partial_{\alpha}\bar{\psi}\:\gamma^3
	\partial_{\beta}\psi \right], 
\end{align}
while the dependence of the fermionic fields in the original world-sheet dimensionless coordinates is simply $\psi = \psi\left(\frac{\sqrt{\lambda}}{2 \pi} \tau, - \frac{\lambda}{2 \pi J} \sigma + J\right)$. Identifying the term multiplying the factor $\bar{\psi}\psi$ with the mass of the theory: 
\begin{equation}
	\label{AAF mass} m = \frac{2 \pi}{\sqrt{\lambda}}, 
\end{equation}
we can assert the mass-dimensions. 

First, we note that $\lambda$, the 't Hooft coupling, is proportional to $l_s^{-2}$, where $l_s$ is the string length, so that the canonical mass-dimension assigned to the coordinates: $[y] = -1$. Thus, as we demand the action to be dimensionless, we conclude from the kinetic and the mass term that: $[\psi] = \frac{1}{2}$ and $[m]=1$. Clearly, this amounts to $[\lambda] = -2$, hence, regarding only the free terms, the mass-dimensions attributed to fields are so far consistent. However, when we turn to the interaction terms of (\ref{AAF S mass action}), we see that in the two-particle interaction term there is one additional mass-dimension, while in the three-particle one, there are two. Therefore, one must introduce in the Lagrangian two coupling constants for the two- and three-particle interaction vertices, and assign to each of them the corresponding dimension: $g_2$, with $[g_2] = -1$; and $g_3$, with $[g_3] = -2$, respectively. The action (\ref{AAF S mass action}), thus, becomes: 
\begin{align}
	\label{AAF action} S &= \int dy^0 \: \int _0^J dy^1 \: \left[ i \bar{\psi} \delslash \psi \: - m \bar{\psi} \psi + \frac{g_2}{4m} \epsilon^{\alpha \beta} \left( \bar{\psi}
	\partial_{\alpha} \psi \; \bar{\psi}\: \gamma^3 
	\partial_{\beta} \psi -
	\partial_{\alpha}\bar{\psi} \psi \; 
	\partial_{\beta} \bar{\psi}\: \gamma^3 \psi \right) \right.- \nonumber \\
	&- \left. \frac{g_3}{16m} \epsilon^{\alpha \beta} \left(\bar{\psi}\psi\right)^2 
	\partial_{\alpha}\bar{\psi}\:\gamma^3
	\partial_{\beta}\psi \right]. 
\end{align}
At this stage the two coupling constants are not independent, since they are derived from the AAF model, containing only one parameter $\lambda$. However, it is interesting to relax this condition, and consider a more general model in which the two coupling constants are independent. This generalization is also convenient for the perturbative analysis, as it allows to keep track of contributions from different vertices. Then, the requirement of quantum integrability, as we will show below, relates the two coupling constants in a manner consistent with the classical dimensionless action (\ref{AAF S mass action}). For the rest of the parameter space our generalized model remains a well-defined quantum field theory, though, non-integrable.

It is important to stress that our action (\ref{AAF action}) essentially differs from the one used by \cite{Klose:2006dd}\footnote{See equation (4.3) of \cite{Klose:2006dd}.} in three aspects. First, there was only one coupling constant introduced in \cite{Klose:2006dd} for both interaction vertices, which is not correct, since $g_2$ and $g_3$ have different dimensions. In addition, as we have already mentioned above, there is an additional factor of $\frac{1}{2}$ in the last interaction term, which, as we will see, is crucial for the quantum integrability. And finally, our action differs from the one in \cite{Klose:2006dd} by an overall sign. This choice has profound consequences in the analysis of the quantum field theory defined by (\ref{AAF action}), as, for instance, the sign of the free Lagrangian determines the role of creation and annihilation operators in the mode expansion of the fields. In particular, we note that the sign choice of \cite{Klose:2006dd} in (4.3) is not consistent with their mode expansion (4.5) and (4.6). Clearly this affects the interplay between the interaction and the free Lagrangian, the more tangible effect of this change being that the model defined by (4.3) is not quantum integrable, since its $S$-matrix fails to factorize. Unfortunately, the analysis of \cite{Klose:2006dd} was not sensitive to this inaccuracy, and as a result the correct $S$-matrix is in fact the inverse of the one obtained in \cite{Klose:2006dd}. This requires further analysis of excited and bound states.

\subsection{Quantization of the free theory}

The next step it to canonically quantize the free theory defined by the action (\ref{AAF action}), in this case, the massive two-dimensional Dirac fermion, satisfying the Dirac equation: 
\begin{equation}
	\label{AAF Dirac equations} i \delslash \psi - m \psi = 0 
\end{equation}
For the free theory the highly non-local Poisson brackets lead, nevertheless, to standard equal-time anticommutation relations: 
\begin{equation}
	\label{AAF equal time anticommutation relations} \left\{ \psi^a(x), \psi^b(x') \right\} = 0, \quad \left\{ {\psi^a}^{\dagger}(x), {\psi^b}^{\dagger}(x') \right\} = 0, \quad \left\{ \psi^a(x), {\psi^b}^{\dagger}(x') \right\} = \delta^{ab} \delta(x-x'). 
\end{equation}
The free quantum Hamiltonian becomes: 
\begin{equation}
	\label{AAF free Hamiltonian} H = \int dx^1\: \left( -i \bar{\psi} \: \gamma^1 
	\partial_1 \psi + m \bar{\psi}\psi \right). 
\end{equation}
and the field expansion takes the form (see appendix {\bf \ref{Appendix Dirac}} for useful definitions): 
\begin{align}
	\psi(x) &= \int \frac{dp_1}{2 \pi} \left[ a(p_1) u(p) e^{-i p\cdot x} + b(-p_1) v(-p) e^{ip\cdot x}\right], \label{AAF dirac plane wave decomposition} \\
	\bar{\psi}(x) &= \int \frac{dp_1}{2 \pi} \left[ a^{\dagger}(p_1) \bar{u}(p) e^{i p\cdot x} + b^{\dagger}(-p_1) \bar{v}(-p) e^{-ip\cdot x}\right] \label{AAF dirac adj plane wave decomposition}, 
\end{align}
where $p_0 = \omega(p)$. Inverting the relations (\ref{AAF dirac plane wave decomposition}) and (\ref{AAF dirac adj plane wave decomposition}), and using (\ref{AAF equal time anticommutation relations}), we obtain the canonical anticommutation relations for the oscillators: 
\begin{equation}
	\label{AAF canonical anticommutation relations for oscillators} \left\{ a(k_1), a^{\dagger}(p_1) \right\} = 2 \pi \delta(k_1-p_1), \quad \left\{ b(-k_1), b^{\dagger}(-p_1) \right\} = 2 \pi \delta(k_1-p_1). 
\end{equation}
The Hamiltonian (\ref{AAF free Hamiltonian}) reads, then: 
\begin{equation}
	\label{AAF S hamiltonian oscillator} H = \int \frac{dp_1}{2 \pi} p_0 \left[ a^{\dagger}(p_1)a(p_1) - b^{\dagger}(p_1)b(p_1) \right]. 
\end{equation}

The fact that the AAF model is a relativistic invariant theory poses a further obstacle: its propagator is not purely retarded. Therefore, one cannot proceed in the standard manner (see, for example, \cite{Klose:2006dd,Thacker:1980ei,Das:2007tb, Melikyan:2008cy}) to compute the $S$-matrix, where this fact was paramount to control the loop corrections, and to calculate the sum of all Feynman diagrams. Nevertheless, we can employ the same technique used in \cite{Bergknoff:1978wr} to overcome this shortcoming. The idea consists in quantizing the theory not with respect to its true ground state, but to a pseudo-vacuum, which, by definition, is the state annihilated by the field operator: 
\begin{equation}
	\label{AAF pseudovacuum} \psi(x)|0\rangle = 0. 
\end{equation}
In this case, all anti-particle levels are left empty, and the $S$-matrix can be computed by the same methods employed for the non-relativistic theories. Finally, from this ``bare'' $S$-matrix one can obtain the Bethe equations, the solution of which enables one to fill back the Dirac sea, and thus, reconstruct the true ground state \cite{Korepin:1997bk}. It is important to bear in mind that the filling of the Dirac sea should change drastically the spectrum and the $S$-matrix. Naturally, (\ref{AAF pseudovacuum}) implies: 
\begin{equation}
	a(p_1)|0\rangle = b(p_1)|0\rangle = 0 \Rightarrow H|0\rangle =0. 
\end{equation}
We define the pseudo-particle states as: 
\begin{equation}
	a^{\dagger}(p_1) |0\rangle = |p\rangle, \quad b^{\dagger}(p_1) |0\rangle = |\tilde{p}\rangle, 
\end{equation}
Then, it is easy to verify that: 
\begin{equation*}
	H|p\rangle = p_0 |p\rangle, \quad H|\tilde{p}\rangle = - p_0 |\tilde{p}\rangle, 
\end{equation*}
hence, the operators $a^{\dagger}(p_1)$ $\big(a(p_1)\big)$ and $b^{\dagger}(p_1)$ $\big(b(p_1)\big)$ create (annihilate) pseudo-particles with momentum $p_1$ and energy $p_0=+\omega(p)$ and $p_0 = -\omega(p)$. The physical vacuum is, therefore, obtained by exciting all the negative energy modes of the pseudo-vacuum.
Finally, we obtain the purely retarded propagator: 
\begin{align}
	\label{AAF propagator} D(x -x') &= \langle0| T \psi(x) \bar{\psi}(x')|0\rangle =\left(i \delslash +m \right) \int \frac{d^2p}{4 \pi^2}\: \frac{i e^{-i p\cdot (x -x')}}{p^2 - m^2+2i\varepsilon p_0}, 
\end{align}
with $d^2p=dp_0\:dp_1$.

\subsection{Scattering of the pseudo-particles in the AAF Model}

In the following, we will be interested in the computation of the two- and three-particle $S$-matrix. For the sake of simplicity we will only consider the scattering between pseudo-particles with positive energy (the excitations over the pseudo-vacuum created by $a^{\dagger}$), since the scattering involving pseudo-particles with negative energy (created by $b^{\dagger}$) can be easily obtained by an analytical continuation to complex rapidities $\theta$, 
\begin{equation*}
	p_0 = m \cosh \theta \quad \textrm{and} \quad p_1 = m \sinh \theta, \quad \textrm{if} \quad \left\{ 
	\begin{array}{lc}
		\theta = \alpha \in \mathbb{R} &\Rightarrow p_0 \in \mathbb{R}_+ \quad \textrm{and} \quad p_1 \in \mathbb{R} \\
		\theta = i \pi -\alpha, \quad \alpha \in \mathbb{R} &\Rightarrow p_0 \in \mathbb{R}_- \quad \textrm{and} \quad p_1 \in \mathbb{R} 
	\end{array}
	\right. . 
\end{equation*}
Therefore, the pseudo-particles with positive energy are appropriately described by real-valued rapidities $\theta = \alpha \in \mathbb{R}$, accordingly the ones with negative energy are naturally parametrized by imaginary rapidities $\theta = i \pi - \alpha, \alpha \in \mathbb{R}$. Since the $S$-matrix is a meromorphic function of the rapidities \cite{Zamolodchikov:1978xm,Dorey:1996gd}, it simultaneously describes the scattering of both types of pseudo-particles. In this case, without any loss of generality, we can take the external states to be composed solely of pseudo-particles in the positive mass-shell:
\begin{equation}
	| \mathbf{p} \rangle = \prod_{i=1}^n a^{\dagger}(p^i_1)|0\rangle, \quad \langle \mathbf{k}| = \langle 0| \prod_{i=1}^n a(k^i_1). 
\end{equation}
\begin{figure}
	\centering 
	\includegraphics[scale=0.65]{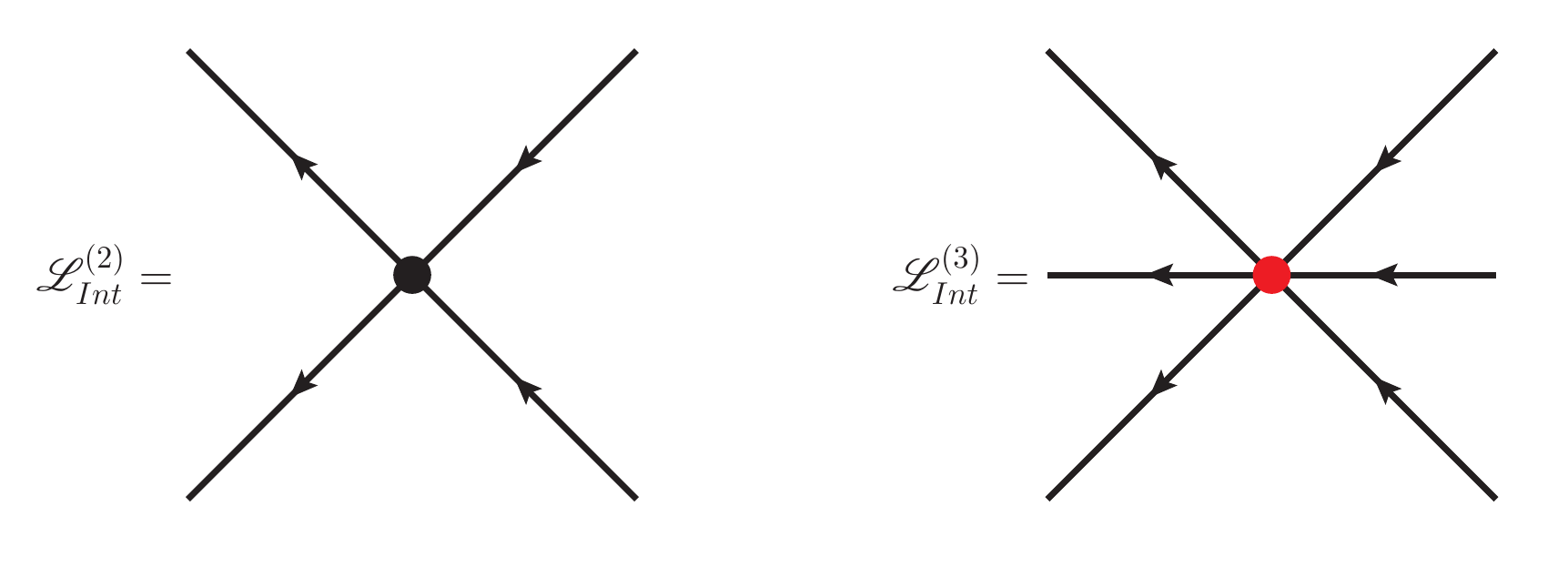} \caption{Interaction vertices in momentum representation.}\label{vertices} 
\end{figure}

For future convenience, we introduce the following notation for the interaction vertices, see figure {\bf \ref{vertices}}, already writing them in normal ordering: 
\begin{align}
	:\mathscr{L}_{Int}^{(2)}: &= \frac{g_2}{4m} \epsilon^{\alpha \beta} :\left( \bar{\psi}
	\partial_{\alpha} \psi \; \bar{\psi}\: \gamma^3 
	\partial_{\beta} \psi -
	\partial_{\alpha}\bar{\psi} \psi \; 
	\partial_{\beta} \bar{\psi}\: \gamma^3 \psi \right): \nonumber \\
	&= - \mathbb{G}^{\alpha \beta}_{ac,bd} \left( \bar{\psi}^a \bar{\psi}^c 
	\partial_{\alpha} \psi^b 
	\partial_{\beta} \psi^d - 
	\partial_{\alpha} \bar{\psi}^a 
	\partial_{\beta} \bar{\psi}^c \psi^b \psi^d \right), \label{AAF int 2 lagrangian} \\
	:\mathscr{L}_{Int}^{(3)}: &= -\frac{g_3}{16m} \epsilon^{\alpha \beta} :\left( \bar{\psi} \psi\right)^2 
	\partial_{\alpha} \bar{\psi} 
	\partial_{\beta} \psi: = - \mathbb{H}^{\alpha \beta}_{ace,bdf} \bar{\psi}^a \bar{\psi}^c 
	\partial_{\alpha} \bar{\psi}^e \psi^b \psi^d 
	\partial_{\beta}\psi^f, \label{AAF int 3 lagrangian} 
\end{align}
where we defined the matrices
\begin{align}
	\mathbb{G}^{\alpha \beta}_{ac,bd} &:= \frac{g_2}{8m} \epsilon^{\alpha \beta} \mathbb{P}_{ac,bd}, \quad \textrm{with} \quad \mathbb{P} := \mathbb{1}_2 \otimes \gamma^3 - \gamma^3 \otimes \mathbb{1}_2, \\
	\mathbb{H}^{\alpha \beta}_{ace,bdf} &:= -\frac{g_3}{16m} \epsilon^{\alpha \beta} \mathbb{Q}_{ace,bdf}, \quad \textrm{with} \quad \mathbb{Q}:= \mathbb{1}_2 \otimes \mathbb{1}_2 \otimes \gamma^3.
\end{align}

\section{Two-particle scattering}\label{2 Particle Scattering}

The two-particle $S$-matrix for the AAF model has been first obtained in \cite{Klose:2006dd}\footnote{See, however, the above discussion on the sign difference.}. Here we give another derivation, based on a general technique which, in principle, can be applied to a large class of integrable models (see, for example, \cite{Das:2007tb, Melikyan:2008cy}). 

\begin{figure}
	[h] \centering 
	\includegraphics[scale=0.75]{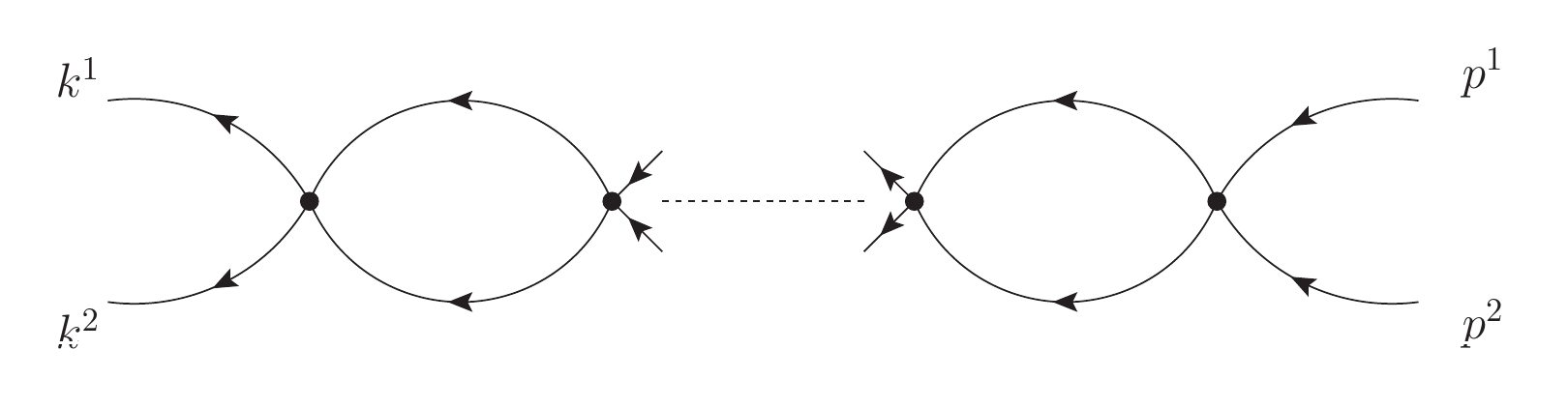} \caption{Bubble diagram for two-pseudo-particle scattering.}\label{bubbles} 
\end{figure}

The fact that the propagator is now purely retarded implies that the two-particle $S$-matrix is given by the sum of bubble diagrams (and its spinorial twists), as depicted in figure {\bf\ref{bubbles}}. Hence, we need only consider the contribution of the two-particle interaction term, $\mathscr{L}^{(2)}_{Int}$. Without any loss in generality, we assume the incoming momenta\footnote{In the following, we will always write the space-time momenta as $p_{\mu}^{i}$, where $\mu =0,1$ is the Lorentz index and $i=1,\ldots, n$, the particle label.} to be ordered: $p_1^1 > p_1^2$ and on-shell.

The two-particle $S$-matrix is then determined from the relation: 
\begin{equation}
	\label{S matrix from scattering amplitude} \langle k^1 k^2 | \hat{S} | p^1 p^2\rangle = S(p^1,p^2) \delta^{(2)}_-(p^1,p^2;k^1,k^2), 
\end{equation}
where: 
\begin{equation}
	\label{delta pm} \delta_{\pm}^{(2)}(p^1,p^2;k^1,k^2) = 4 \pi^2 \left[ \delta(k^1_1 - p^1_1) \delta(k^2_1-p^2_1) \pm \delta(k^1_1 - p^2_1) \delta(k^2_1-p^1_1) \right], 
\end{equation}
and the scattering amplitude is given by: 
\begin{align}
	\langle k^1 k^2 | \hat{S} | p^1 p^2\rangle &= \langle k^1 k^2 | e^{i \int \mathscr{L}^{(2)}_{Int}\: d^2x} | p^1 p^2\rangle \nonumber \\
	&= \langle k^1 k^2 | p^1 p^2\rangle + i \langle k^1 k^2 | \int \mathscr{L}^{(2)}_{Int} \: d^2x \: | p^1 p^2\rangle \:- \nonumber \\
	&- \frac{1}{2} \langle k^1 k^2 | T \left( \int \mathscr{L}^{(2)}_{Int} \: d^2x \right)^2| p^1 p^2\rangle + \cdots . 
\end{align}
The non-scattering term is easily computed and yields: 
\begin{equation}
	\label{2p non-scattering term} \langle k^1 k^2 | p^1 p^2\rangle = \delta^{(2)}_-(p^1,p^2;k^1,k^2). 
\end{equation}
At the tree level, we need to evaluate: 
\begin{align}
	\langle k^1 k^2 | \hat{S} | p^1 p^2\rangle \Big|_{g_2} &= i \langle k^1 k^2 |  \int \mathscr{L}^{(2)}_{Int} \: d^2x\: | p^1 p^2\rangle \nonumber \\
	&= -i \mathbb{G}^{\alpha \beta}_{ac,bd} \langle k^1 k^2 | \int d^2x \left( \bar{\psi}^a \bar{\psi}^c 
	\partial_{\alpha} \psi^b 
	\partial_{\beta} \psi^d - 
	\partial_{\alpha} \bar{\psi}^a 
	\partial_{\beta} \bar{\psi}^c \psi^b \psi^d \right)| p^1 p^2\rangle. 
\end{align}
We can compute all the integrals to obtain the off-shell\footnote{To be more precise, the outcoming pseudo-particles are off-shell, though the assumption of on-shell incoming pseudo-particles is not used in the derivation of (\ref{2p tree-level off-shell}), thus justifying calling it off-shell.} tree level term: 
\begin{align}
	\label{2p tree-level off-shell} \langle k^1 k^2 | \hat{S} | p^1 p^2\rangle \Big|_{g_2} &= -\frac{ig_2}{8m} \left\{ k^2 \times k^1 \left( \bar{U}^k_{12}+ \bar{U}^k_{21} \right) \mathbb{P} \: (U^p_{12} - U^p_{21}) \right. \nonumber \\
	&+ \left. p^2 \times p^1 \left( - \bar{U}^k_{12}+ \bar{U}^k_{21} \right) \mathbb{P} \: (U^p_{12} + U^p_{21}) \right\} 4 \pi^2 \delta^{(2)}\left(k^1 +k^2 -p^1-p^2\right). 
\end{align}
where we introduced the shorthand notation for the bi-spinors: 
\begin{equation*}
	\bar{U}^k_{ij} = \bar{u}(k^i) \otimes \bar{u}(k^j), \quad U^p_{ij} = u(p^i) \otimes u(p^j), 
\end{equation*}
and denoted: $\epsilon^{\alpha \beta} p^i_{\alpha} p^j_{\beta} = p^i \times p^j$.

Taking the outcoming pseudo-particles to be on-shell as well, we can use the identity: 
\begin{equation}
	\label{on-shell relation for delta} 4 \pi^2 \delta^{(2)}\left(k^1 +k^2 -p^1-p^2\right) = \left| \frac{p^1_0 \: p^2_0}{p^2 \times p^1}\right| \delta^{(2)}_+(p^1,p^2;k^1,k^2), 
\end{equation}
together with the fact that for our ordering of incoming momenta $p^2 \times p^1 >0$ to write: 
\begin{equation}
	\langle k^1 k^2 | \hat{S} | p^1 p^2\rangle \Big|_{g_2} = -\frac{ig_2}{2m} \: p^1_0 \: p^2_0 \: \bar{U}^p_{21} \: \mathbb{P} \: U^p_{12} \: \delta^{(2)}_-(p^1,p^2;k^1,k^2). 
\end{equation}
Finally, we compute the spinorial product 
\begin{equation}
	\label{2p tree level spinorial product} \bar{U}^p_{21} \: \mathbb{P} \: U^p_{12} = \frac{p^2 \times p^1}{p^1_0 \: p^2_0}, 
\end{equation}
to obtain the on-shell tree-level amplitude: 
\begin{equation}
	\label{2p tree-level on-shell} \langle k^1 k^2 | \hat{S} | p^1 p^2\rangle \Big|_{g_2} = -\frac{ig_2}{2m} \left( p^2 \times p^1 \right) \delta^{(2)}_-(p^1,p^2;k^1,k^2). 
\end{equation}

The one-loop amplitude computation is rather more involved as it contains more terms to take into account. Namely, we have to evaluate: 
\begin{align}
	\langle k^1 k^2 | \hat{S} | p^1 p^2\rangle \Big|_{g_2^2} &= - \frac{1}{2} \langle k^1 k^2 | T \left( \int \mathscr{L}^{(2)}_{Int} \: d^2x \right)^2| p^1 p^2\rangle \nonumber \\
	&= - \frac{1}{2} \mathbb{G}^{\alpha \beta}_{ac,bd} \mathbb{G}^{\gamma \delta}_{eg,fh} \langle k^1 k^2 | T \int d^2x \: d^2y\: \left[ \bar{\psi}^a \bar{\psi}^c 
	\partial_{\alpha} \psi^b 
	\partial_{\beta} \psi^d \bar{\phi}^e \bar{\phi}^g 
	\partial_{\gamma} \phi^f 
	\partial_{\delta} \phi^h - \right. \nonumber \\
	&- \left. \bar{\psi}^a \bar{\psi}^c 
	\partial_{\alpha} \psi^b 
	\partial_{\beta} \psi^d 
	\partial_{\gamma} \bar{\phi}^e 
	\partial_{\delta} \bar{\phi}^g \phi^f \phi^h - 
	\partial_{\alpha} \bar{\psi}^a 
	\partial_{\beta} \bar{\psi}^c \psi^b \psi^d \bar{\phi}^e \bar{\phi}^g 
	\partial_{\gamma} \phi^f 
	\partial_{\delta} \phi^h + \right. \nonumber \\
	&+ \left. 
	\partial_{\alpha} \bar{\psi}^a 
	\partial_{\beta} \bar{\psi}^c \psi^b \psi^d 
	\partial_{\gamma} \bar{\phi}^e 
	\partial_{\delta} \bar{\phi}^g \phi^f \phi^h \right]| p^1 p^2\rangle, 
\end{align}
where to avoid cluttering, we denoted $\psi \equiv \psi(x)$ and $\phi \equiv \psi(y)$. Expanding the $T$-product, we realize that only the terms with two contractions of the type 
\begin{equation*}
	\overbracket[0.5pt]{\psi \bar{\phi}}\: \overbracket[0.5pt]{\psi \bar{\phi}}\:, \quad \textrm{or} \quad \overbracket[0.5pt]{\bar{\psi} \phi}\: \overbracket[0.5pt]{\bar{\psi} \phi} 
\end{equation*}
contribute. The other terms vanish identically, either because the inner product 
\begin{equation*}
\langle k^1 k^2| f(a, a^{\dagger})| p^1 p^2\rangle = 0 \quad \text{or} \quad \langle k^1 k^2| g(\psi, \bar{\psi})| p^1  p^2\rangle \propto \theta(x^0-y^0) \theta(y^0 - x^0)= 0, 
\end{equation*}
with $f(a, a^{\dagger})$ and $ g(\psi, \bar{\psi})$ being arbitrary functions. Hence, 
\begin{align}
	\label{2p 1-loop} \langle k^1 k^2 | \hat{S} | p^1 p^2\rangle &\Big|_{g_2^2} = 2 \left( \frac{g_2}{8m}\right)^2 \left[ \left(k^2 \times k^1 \right) \left(p^2 \times p^1 \right) \left(\bar{U}^{k}_{12} + \bar{U}^{k}_{21}\right) \mathbb{P}\: I_0(p^1,p^2) \: \mathbb{P} \left( U^p_{12} + U^p_{21}\right) - \right. \nonumber \\
	&- \left. \left(p^2 \times p^1 \right) \left(\bar{U}^{k}_{12} - \bar{U}^{k}_{21}\right) \mathbb{P} \: I_1(p^1,p^2) \: \mathbb{P} \left( U^p_{12} + U^p_{21}\right) - \right. \nonumber \\
	&- \left. \left(k^2 \times k^1 \right) \left(\bar{U}^{k}_{12} + \bar{U}^{k}_{21}\right) \mathbb{P} \: I_1(p^1,p^2) \: \mathbb{P} \left( U^p_{12} - U^p_{21}\right) + \right. \nonumber \\
	&+ \left. \left(\bar{U}^{k}_{12} - \bar{U}^{k}_{21}\right) \mathbb{P} \: I_2(p^1,p^2) \: \mathbb{P} \left( U^p_{12} - U^p_{21}\right) \right] 4\pi^2 \delta^{(2)}\left(k^1 +k^2 -p^1-p^2\right), 
\end{align}
where $I_i(p^1,p^2)$, $i=0,1,2$ are defined in appendix {\bf\ref{Integrals}} by equations (\ref{I_0}), (\ref{I_1}) and (\ref{I_2}).

The completeness relations (\ref{appendix dirac 2d completeness relations}), together with our ordering for the incoming momenta and the identity for the on-shell momenta: 
\begin{equation}
	\label{relation for on-shell momenta} \frac{\left( p^2 \times p^1 \right) \left( p^1_0 + p^2_0 \right)}{\left( p^1_1 - p^2_1 \right) \left( p^1 + p^2 \right)^2} = \frac{1}{2}, 
\end{equation}
lead to the following central relation between the one-loop amplitude and the off-shell\footnote{Here we actually mean that only the outcoming pseudo-particles are off-shell.} tree-level amplitude (\ref{2p tree-level off-shell}) 
\begin{equation}
	\label{2p 1-loop off-shell} \langle k^1 k^2 | \hat{S} | p^1 p^2\rangle \Big|_{g_2^2} = 2 \left(p^2 \times p^1\right) \left( \frac{-i g_2}{8m} \right) \langle k^1 k^2 | \hat{S} | p^1 p^2\rangle \Big|_{g_2} . 
\end{equation}
Since the outcoming pseudo-particles are off-shell, equation (\ref{2p 1-loop off-shell}) suggests that we can regard the one-loop scattering amplitude as the interaction vertex in momentum representation, with the incoming momenta $p^1$ and $p^2$ on-shell, multiplied by some function of this pair of momenta and the coupling constant. Therefore the $n$-loop scattering amplitude corresponds to the product of $n$ of these modified vertices: 
\begin{align}
	\langle k^1 k^2 | \hat{S} | p^1 p^2\rangle \Big|_{g_2^{n+1}} &= \left[ 2 \left(p^2 \times p^1\right) \left( \frac{-i g_2}{8m} \right) \right]^{n} \langle k^1 k^2 | \hat{S} | p^1 p^2\rangle \Big|_{g_2}  \nonumber \\
	&\stackrel{on-shell}{=} 2 \left( -\frac{ig_2}{4m}\left(p^2 \times p^1\right) \right)^{n+1} \delta_-(p^1,p^2;k^1,k^2), 
\end{align}
where in the last step, we took the outcoming pseudo-particles $k^1$ and $k^2$ on-shell.

We are now in the position to obtain the full scattering amplitude: 
\begin{align}
	\langle k^1 k^2 | \hat{S} | p^1 p^2\rangle &= \langle k^1 k^2 | p^1 p^2\rangle + \sum_{n=1}^{\infty} \langle k^1 k^2 | \hat{S} | p^1 p^2\rangle \Big|_{g_2^n} \nonumber \\
	&= \frac{1 - \frac{ig_2}{4m} p^2 \times p^1}{1 + \frac{ig_2}{4m} p^2 \times p^1} \delta_-(p^1,p^2;k^1,k^2), 
\end{align}
from which, by comparison with (\ref{S matrix from scattering amplitude}), we read off the $S$-matrix for the scattering of two-particle, 
\begin{equation}
	\label{2p S matrix} S(p^1,p^2) = \frac{1 - \frac{ig_2}{4m} p^2 \times p^1}{1 + \frac{ig_2}{4m} p^2 \times p^1}. 
\end{equation}
It is crucial to notice that our $S$-matrix is the inverse of the one derived by \cite{Klose:2006dd}.

\section{Three-particle scattering}\label{3 Particle Scattering}

In this section we analyze the $S$-matrix factorization, which reflects the quantum integrability of the model. The first step in this program is to consider the $S$-matrix for the scattering of three pseudo-particles and confirm that it can be properly written as the product of three $S$-matrices for two-particle scattering (\ref{2p S matrix}). Before going into the details of the actual three-particle scattering amplitudes, just as we did for the two-particle case in the previous section, we take a closer look at the factorizable expression for the three-particle $S$-matrix, namely, 
\begin{align}
	\label{3p S matrix from 2p S matrix} S(p^1,p^2,p^3) &= S(p^1,p^2)S(p^1,p^3)S(p^2,p^3) \nonumber \\
	&=\frac{1-\frac{ig_2}{4m} p^2 \times p^1}{1 + \frac{ig_2}{4m} p^2 \times p^1} \cdot \frac{1-\frac{ig_2}{4m} p^3 \times p^1}{1 + \frac{ig_2}{4m} p^3 \times p^1} \cdot \frac{1-\frac{ig_2}{4m} p^3 \times p^2}{1 + \frac{ig_2}{4m} p^3 \times p^2} \nonumber \\
	&= 1 + 2 \sum_{n=1}^3 \left[ \left(- \frac{ig_2}{4m}\right) \left( p^2 \times p^1 + p^3 \times p^1 + p^3 \times p^2 \right)\right]^n + \nonumber \\
	&+ 2\left(\frac{ig_2}{4m} \right)^3 \left(p^2\times p^1 + p^3 \times p^1\right) \left(p^2\times p^1 + p^3 \times p^2\right) \left(p^3\times p^1 + p^3 \times p^2 \right) + \nonumber \\
	&+ O(g_2^4). 
\end{align}
In the following, we will compute the three-particle scattering amplitude and show that this necessary condition (\ref{3p S matrix from 2p S matrix}) is satisfied, up to the first non-trivial order in $g_2$ and $g_3$, provided a very precise relation between $g_2$ and $g_3$.

\subsection{Diagrammatic calculations}

In this case, we must consider the full interaction Lagrangian $\mathscr{L}_{Int} = \mathscr{L}_{Int}^{(2)} + \mathscr{L}_{Int}^{(3)}$, as both the initial and final states involve three pseudo-particles: 
\begin{equation*}
	|\mathbf{p}\rangle = |p^1p^2p^3\rangle = a^{\dagger}\left(p^1_1\right)a^{\dagger}\left(p^2_1\right)a^{\dagger}\left(p^3_1\right)|0\rangle \quad \textrm{and} \quad \langle \mathbf{k}| = \langle k^1k^2k^3|=\langle 0| a\left(k^3_1\right)a\left(k^2_1\right)a \left(k^1_1\right). 
\end{equation*}
Similarly, we will assume without any loss of generality that the incoming pseudo-particles are on-shell and have their momenta ordered: $p_1^1 > p_1^2 > p_1^3$. The analyticity in the coupling constants of the three-particle scattering amplitude implies: 
\begin{equation}
	\langle \mathbf{k}| \hat{S} | \mathbf{p} \rangle = \langle \mathbf{k}| \mathbf{p} \rangle + \langle \mathbf{k}| \hat{S} | \mathbf{p} \rangle \Big|_{g} + \langle \mathbf{k}| \hat{S} | \mathbf{p} \rangle \Big|_{g^2} + \cdots, 
\end{equation}
where $g$ stands either for $g_1$ or $g_2$. The non-scattering term is easily computed: 
\begin{equation}
	\label{3p non-scattering term} \langle \mathbf{k}| \mathbf{p}\rangle = 3!(2 \pi)^3 \mathcal{A}_{p}\left[ \delta(k^1-p^1) \delta(k^2-p^2) \delta(k^3-p^3) \right]. 
\end{equation}
Here, we introduced the antisymmetrization operator, or simply, antisymmetrizator, defined by: 
\begin{equation}
	\mathcal{A}_q[f(\mathbf{q})] := \frac{1}{3!} \sum_A \mathrm{sign}(A) f(A[\mathbf{q}]), 
\end{equation}
with the sum taken over all possible permutations of $(1,2,3)$ and the vector $A[\mathbf{q}] := \left(q^{A_1},q^{A_2},q^{A_3} \right)$.

For the tree-level amplitude we need to evaluate: 
\begin{align}
	\langle \mathbf{k}| \hat{S} | \mathbf{p} \rangle \Big|_{g} &= i \langle \mathbf{k}|T \int d^2x \: \left( \mathscr{L}_{Int}^{(2)} + \mathscr{L}_{Int}^{(3)} \right) | \mathbf{p} \rangle, 
\end{align}
leading to the off-shel tree-level amplitude, 
\begin{align}
	\label{3p tree-level off-shell} \langle \mathbf{k}| \hat{S} | \mathbf{p} \rangle \Big|_{g} &= (3!)^2 \mathcal{A}_{p,k} \left[\frac{ig_2}{8m} \left( k^2\times k^1 + p^2\times p^1 \right) \; \bar{U}^k_{12} \; \mathbb{P} \; U^p_{21} \; 2 \pi \delta(k^3-p^3) \: - \right. \nonumber \\
	&- \left. \frac{ig_3}{16m} \left(k^3\times p^3 \right) \bar{U}^k_{123} \; \mathbb{Q} \; U^p_{123} \right] \; 4 \pi^2 \delta^{(2)}(\mathbf{k}-\mathbf{p}), 
\end{align}
where we introduced the almost self-evident notation for the tri-spinors: 
\begin{equation*}
	\bar{U}^k_{ijl} = \bar{u}(k^i) \otimes \bar{u}(k^j) \otimes \bar{u}(k^l), \quad U^p_{ijl} = u(p^i) \otimes u(p^j) \otimes u(p^l). 
\end{equation*}
Proceeding in the same way as with the two-particle case, we can impose the mass-shell condition on the outcoming pseudo-particles in the first term from (\ref{3p tree-level off-shell}), so that we can apply the identity (\ref{on-shell relation for delta}) and use the fact that for our ordering of initial momenta $p^i \times p^j >0$ if $i>j$ together with (\ref{2p tree level spinorial product}) to obtain: 
\begin{align}
	\label{3p tree-level on-shell} \langle \mathbf{k}| \hat{S} | \mathbf{p} \rangle \Big|_{g} &= -\frac{i g_2}{2m} \left[ p^2 \times p^1 + p^3 \times p^1 + p^3 \times p^2 \right] \langle \mathbf{k}| \mathbf{p}\rangle \:- \nonumber \\
	&- \frac{ig_3}{16m} (3!)^2 \mathcal{A}_{p,k} \left[ k^3 \times p^3 \: \bar{U}^k_{123} \: \mathbb{Q} \: U^p_{123} \right] 4 \pi^2 \delta^{(2)}(\mathbf{k}-\mathbf{p}). 
\end{align}

It is crucial to realize that there is no identity similar to (\ref{on-shell relation for delta}) involving the three momenta. Hence, it is not possible to compute the spinorial product within the second term in (\ref{3p tree-level off-shell}), so as to reduce it to an expression proportional to $\langle \mathbf{k} | \mathbf{p} \rangle$. Remembering that for an integrable model the $S$-matrix is expected to be of the form: 
\begin{equation}
	\langle \mathbf{k}| \hat{S} | \mathbf{p} \rangle = S(\mathbf{p}) \langle \mathbf{k}| \mathbf{p} \rangle \label{smat_form}, 
\end{equation}
it is then clear that not only one cannot derive the expression (\ref{3p S matrix from 2p S matrix}) in the tree-level approximation, but it is not even possible to write the $S$-matrix in the form (\ref{smat_form}). Even though this seems to be a formidable obstacle to prove quantum integrability, it is not so. Indeed, the same situation arises in a similar calculation for the Landau-Lifshitz model \cite{Melikyan:2008cy}, where it was found that such troublesome terms\footnote{Such as the second term in (\ref{3p tree-level off-shell}).} are cancelled out by certain contributions coming from higher order (in $g$) scattering amplitudes. 
\begin{figure}
	[t] \centering 
	\includegraphics[scale=0.75]{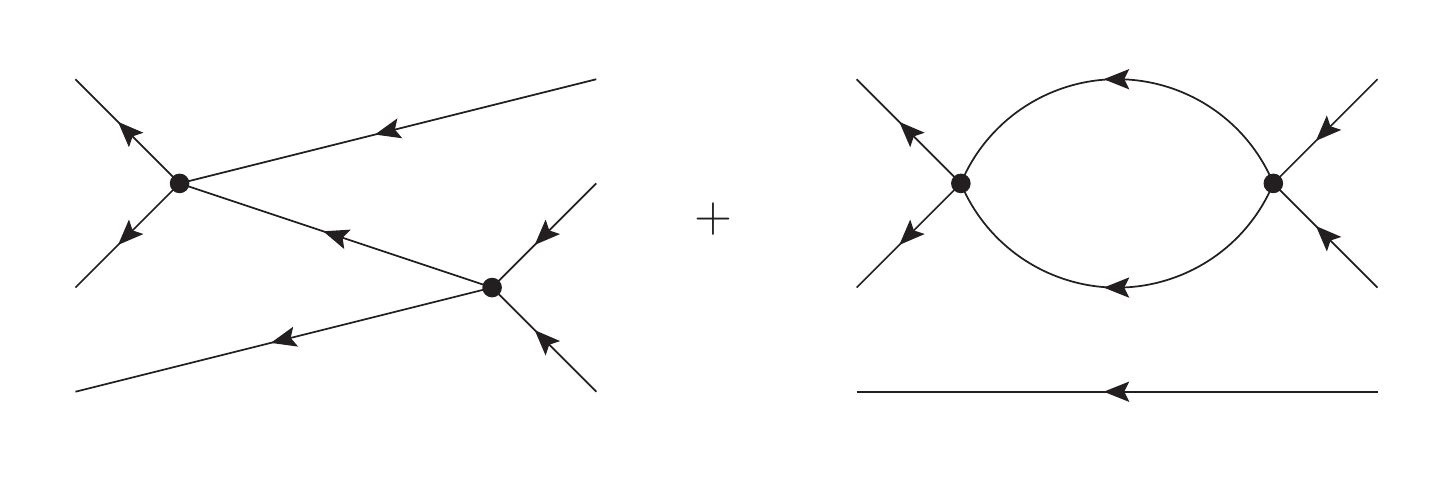} \caption{Feynman diagrams for the one-loop scattering amplitude $V_1^{(2)} \left( \mathbf{k},\mathbf{p} \right)$. Here, the first graph corresponds to the principal value contribution coming from the one-contraction terms, while the second corresponds to the sum of the delta contribution from the one-contraction term with the contribution coming from the term with two contractions.}\label{2ndorder22} 
\end{figure}
\begin{figure}
	[t] \centering 
	\includegraphics[scale=0.75]{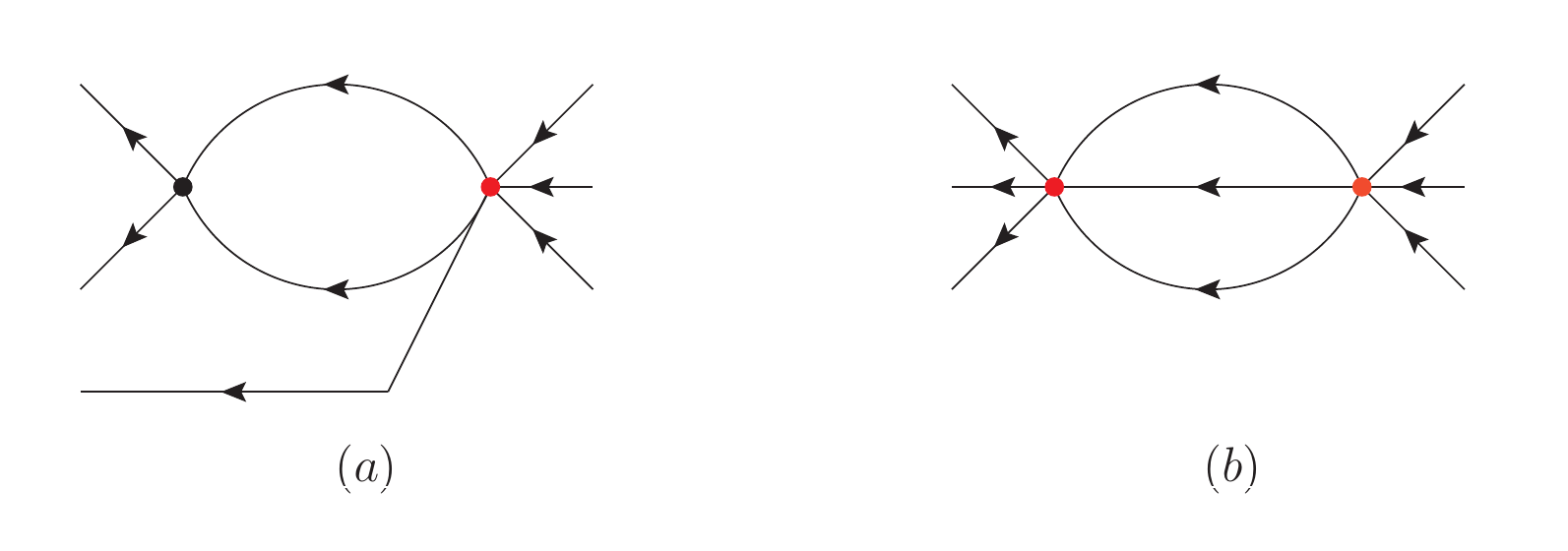} \caption{Feynman diagrams corresponding to the one-loop amplitudes: $(a)$ $V_2^{(2)} \left( \mathbf{k},\mathbf{p} \right)$ and $V_3^{(2)} \left( \mathbf{k},\mathbf{p} \right)$, $(b)$ $V_4^{(2)} \left( \mathbf{k},\mathbf{p} \right)$.}\label{2ndorder23} 
\end{figure}

We consider here the one-loop scattering amplitude to track down the contribution to cancel out the second term in (\ref{3p tree-level off-shell}) and, thus, render the tree-level $S$-matrix factorizable. 
\begin{align}
	\label{3p 1-loop expansion} \langle \mathbf{k}| \hat{S} | \mathbf{p} \rangle \Big|_{g^2} &= -\frac{1}{2} \langle \mathbf{k}|T\left[ \int d^2 x \: \left(\mathscr{L}^{(2)}_{Int} + \mathscr{L}^{(3)}_{Int}\right)\right]^2| \mathbf{p}\rangle \notag \\
	&= -\frac{1}{2} \langle \mathbf{k}| T \int d^2x\: d^2y\: \left[ \mathscr{L}^{(2)}_{Int}(x)\mathscr{L}^{(2)}_{Int}(y) + \mathscr{L}^{(2)}_{Int}(x)\mathscr{L}^{(3)}_{Int}(y) + \mathscr{L}^{(3)}_{Int}(x)\mathscr{L}^{(2)}_{Int}(y) \: + \right. \notag \\
	&+ \left. \mathscr{L}^{(3)}_{Int}(x)\mathscr{L}^{(3)}_{Int}(y) \right] | \mathbf{p}\rangle \notag \\
	&\equiv V^{(2)}_1(\mathbf{k},\mathbf{p}) + V^{(2)}_2(\mathbf{k},\mathbf{p}) +V^{(2)}_3(\mathbf{k},\mathbf{p}) + V^{(2)}_4(\mathbf{k},\mathbf{p}). 
\end{align}
As noted in \cite{Melikyan:2008cy}, this cancellation can only happen between the diagrams of the same order in $\hbar$.\footnote{We consider here the loop expansion which corresponds to an expansion in powers of $\hbar$. Thus, it is clear that diagrams with a different number of loops cannot cancel each other. Nevertheless, we stress that there are two different coupling constants $g_2$ and $g_3$, the dimensions of which are such that the second diagram of figure \ref{vertices} and the first diagram of figure \ref{2ndorder22} are of the same order, and may cancel each other (we show below that it is indeed the case). Therefore, one must consider all possible diagrams for a given order in $\hbar^n$.} Clearly, all tree-level diagrams are of the order $\hbar^0$, hence, at this stage, we can focus only on the first term from the expansion (\ref{3p 1-loop expansion}), $V^{(2)}_1(\mathbf{k},\mathbf{p})$, as it is the only term which contains diagrams of the zeroth order in $\hbar$. The corresponding Feynman diagrams are depicted in figures {\bf \ref{2ndorder22}} and {\bf \ref{2ndorder23}}.

The analysis of the time-ordered product expansion goes along the same lines as in the two-particle case, since only the two-particle interaction term comes into play in $V^{(2)}_1$. However, as there are now three pseudo-particles in the initial and final states, the contribution of the terms with only one contraction becomes also non-zero. Therefore, one must take into account the contributions coming from the terms containing only the following contractions: 
\begin{equation*}
	\overbracket[0.5pt]{\psi \bar{\phi}}\:, \: \overbracket[0.5pt]{\bar{\psi}\phi}\:, \: \overbracket[0.5pt]{\psi \bar{\phi}}\: \overbracket[0.5pt]{\psi \bar{\phi}}\:, \: \textrm{or} \: \overbracket[0.5pt]{\bar{\psi} \phi}\: \overbracket[0.5pt]{\bar{\psi} \phi}. 
\end{equation*}
The contribution of the two-contraction terms can be written in the form: 
\begin{align}
	\label{3p 1-loop 2-contraction off-shell} V^{(2)}_{1(2C)}(\mathbf{k},\mathbf{p}) &= \frac{1}{2} \left( - \frac{3! \:g_2}{4m} \right)^2 \mathcal{A}_{p,k} \left\{ \bar{U}_{12}^k \: \mathbb{P} \left[ \left( p^2 \times p^1 \right) \left( k^2 \times k^1 \right) I_0(p^1,p^2) + \left( k^2 \times k^1 - \right. \right. \right. \notag \\
	&- \left. \left. \left. p^2 \times p^1 \right) I_1(p^1,p^2) - I_2(p^1,p^2) \right] \mathbb{P} \: U^p_{21} \: 8 \pi^3 \delta^{(2)}(k^1+k^2-p^1-p^2) \delta(k^3 - p^3) \right\},
\end{align}
with the integrals $ I_0(p^1,p^2) $, $ I_1(p^1,p^2) $ and $ I_2(p^1,p^2) $ defined in appendix {\bf\ref{Integrals}}. 
Again, by using the completeness relations (\ref{appendix dirac 2d completeness relations}) and imposing the mass-shell condition, we can employ the identity (\ref{relation for on-shell momenta}) in conjunction with our ordering for the incoming momenta to conclude that 
\begin{equation}
	\label{3p 1-loop 2-contraction on-shell} V^{(2)}_{1(2C)}(\mathbf{k},\mathbf{p}) = 2 \left(-\frac{i g_2}{4m}\right)^2 \left[ (p^2\times p^1)^2 + (p^3\times p^1)^2 + (p^3\times p^2)^2\right] \langle \mathbf{k}| \mathbf{p}\rangle. 
\end{equation}

On the other hand, the evaluation of the one-contraction terms is considerably more complex, and can be written in the form: 
\begin{align}
	\label{3p 1-loop 1-contraction off-shell} V^{(2)}_{1(1C)}(\mathbf{k},\mathbf{p}) &= \left(- \frac{3! \: g_2}{4 m} \right)^2 \mathbb{P}_{ac,bd} \: \mathbb{P}_{eg,fh} \: \mathcal{A}_{p,k} \bigg\{ \Big[ -\left( p^1 \times p^2 \right) \left( k^2 \times k^3 \right) I + \epsilon^{\alpha \beta} \Big( \left( p^1 \times p^2 \right) \: + \Big. \Big. \bigg. \notag \\
	&+ \bigg. \Big. \Big. \left( k^2 \times k^3 \Big) k^1_{\beta} \right) I_{\alpha} - \epsilon^{\alpha \beta} \epsilon^{\gamma \delta} k^1_{\beta} \: p^3_{\delta} \: I_{\alpha \gamma} \Big]^{fa} \bar{u}^c(k^1_1) \bar{u}^e(k^2_1) \bar{u}^g(k^3_1) u^b(p^1_1) u^d(p^2_1) u^h(p^3_1) \bigg\}, 
\end{align}
where the integrals $I$, $I_{\alpha}$ and $I_{\alpha \gamma}$ are defined in appendix {\bf \ref{Integrals}} by equations (\ref{3p I}), (\ref{3p I alpha}) and (\ref{3p I alpha gamma}), respectively. One crucial feature of the aforementioned integrals is that they all are proportional to a sum of the delta term and the principal value (\rm{p.v.}) term: 
\begin{equation}
	\label{delta + pv split} \underbrace{\frac{ \dslash{\Delta} + m }{4 \omega(\Delta)} \: 2 \pi \Big[ \delta \left( \Delta_0 - \omega(\Delta) \right) - \delta \left( \Delta_0 + \omega(\Delta) \right) \Big]}_{\textrm{delta term}} + \underbrace{\frac{i\left( \dslash{\Delta} + m \right)}{\Delta^2 - m^2}}_{\textrm{p.v. term}},
\end{equation}
where we have defined $\Delta \equiv p^1 + p^2 - k^1$.

This split of the one-contraction contribution plays an important role in the subsequent analysis. As we will see, the contribution of the delta term from (\ref{delta + pv split}) to (\ref{3p 1-loop 1-contraction off-shell}) will combine with (\ref{3p 1-loop 2-contraction on-shell}) to yield the factorizable $S$-matrix at one-loop, while the one coming from the \rm{p.v.} term in (\ref{delta + pv split}) will cancel the contribution of the three-particle interaction Lagrangian at tree-level, which prevented $S$ matrix factorability. Upon substitution of the integrals $I$, $I_{\alpha}$ and $I_{\alpha \gamma}$, the expression (\ref{3p 1-loop 1-contraction off-shell}) greatly simplifies, 
\begin{align}
	\label{3p 1-loop 1-contraction off-shell 2} V^{(2)}_{1(1C)}(\mathbf{k},\mathbf{p}) &= \left(- \frac{3! \:i g_2}{4 m} \right)^2 \mathcal{A}_{p,k} \Bigg\{ \left( p^2 \times p^1 + \Delta \times k^1 \right) \left( k^3 \times k^2 + \Delta \times p^3 \right) \Bigg. \left[ \frac{i}{\Delta^2 - m^2} \: + \right. \nonumber \\
	&+ \Bigg. \left. \frac{2 \pi}{4 \omega(\Delta)} \Big( \delta \left( \Delta_0 - \omega(\Delta) \right) - \delta \left( \Delta_0 + \omega(\Delta) \right) \Big) \right] \bar{U}^k_{123} \: \mathbb{M} (\Delta) \: U^p_{213} \Bigg\} \: 4 \pi^2 \delta^{(2)}\left( \mathbf{k} - \mathbf{p} \right), 
\end{align}
where we introduced 
\begin{equation}
	\label{M matrix} \mathbb{M}(q) := \gamma^3 \otimes (\dslash{q}+m)\otimes \gamma^3 - \gamma^3 \otimes \gamma^3 (\dslash{q}+m) \otimes \mathbb{1}_2 - \mathbb{1}_2 \otimes (\dslash{q}+m)\gamma^3 \otimes \gamma^3 + \mathbb{1}_2 \otimes \gamma^3 (\dslash{q}+m) \gamma^3 \otimes \mathbb{1}_2 . 
\end{equation}

The contribution of the delta term is easier to evaluate, as the delta functions implement the mass-shell conditions for the pseudo-particles with positive and negative energies, respectively. Since we assumed the scattering pseudo-particles to have positive energy, the condition $\Delta_0 + \omega(\Delta)$ cannot be satisfied for any $\mathbf{p}$ and $\mathbf{k}$, and therefore, we can disregard the second delta function above. We can conveniently rewrite the positive-energy delta function as follows: 
\begin{equation}
	\delta \Big( \Delta_0 - \omega(\Delta) \Big) = \frac{p^1_0 \: p^2_0}{|p^2 \times p^1|} \Big[ \delta(k^1_1 - p^1_1) + \delta(k^1_1 - p^2_1) \Big], 
\end{equation}
so that the delta function for overall energy-momentum conservation can be further simplified, allowing us to use the identity (\ref{on-shell relation for delta}) and our ordering of initial momenta to trivially compute the spinorial products, along the same lines as with the two-particle case. After long but straightforward calculations the delta term contribution reduces to: 
\begin{equation}
	\label{3p 1-loop 1-contraction delta term} 4 \left( - \frac{i g_2}{4 m}\right)^2 \left[(p^2\times p^1)(p^3 \times p^1) + (p^2\times p^1)(p^3 \times p^2) +(p^3\times p^2)(p^3 \times p^1)\right] \langle \mathbf{k}| \mathbf{p}\rangle . 
\end{equation}
Equations (\ref{3p 1-loop 2-contraction on-shell}) and (\ref{3p 1-loop 1-contraction delta term}) can be easily combined, yielding: 
\begin{align}
	\label{3p 1-loop} V^{(2)}_{1}(\mathbf{k},\mathbf{p}) &= 2 \left(-\frac{i g_2}{4 m}\right)^2 \left[(p^2\times p^1) + (p^3 \times p^2) +(p^3 \times p^1)\right]^2 \langle \mathbf{k}| \mathbf{p}\rangle + \: \notag \\
	&+ i \left(-\frac{3!\: ig_2}{4m}\right)^2 \mathcal{A}_{k,p} \left\{ \frac{\left[p^2\times p^1+\Delta\times k^1\right] \left[k^3 \times k^2 + \Delta \times p^3\right]}{\Delta^2 -m^2} \bar{U}^k_{123}\mathbb{M}(\Delta)U^p_{213}\right\} \cdot \nonumber \\
	&\cdot 4 \pi^2 \delta^{(2)}(\mathbf{k}-\mathbf{p}). 
\end{align}
Clearly, the first term in (\ref{3p 1-loop}) amounts to the complete contribution to a factorizable $S$-matrix at one-loop (\ref{3p S matrix from 2p S matrix}). It is worth pointing out that, according to the scheme proposed in \cite{Melikyan:2008cy}, the remaining terms appearing in the one-loop scattering amplitude (\ref{3p 1-loop expansion}), $V^{(2)}_i(\mathbf{k},\mathbf{p})$, $i=2,3,4$, must be cancelled out by higher order contributions so as to have a factorizable $S$-matrix. 

\subsection{Continuity of the scattering amplitudes} \label{continuity}

Before moving onto the proof of the $S$-matrix factorization at first order, we pause to address one important subtlety that arises during the evaluation of the integrals $I$, $I_{\alpha}$ and $I_{\alpha \gamma}$ (see (\ref{3p 1-loop 1-contraction off-shell})), and which is intimately related to the formal proof of the $S$-matrix factorization for quantum integrable models \cite{Parke:1980ki, Zamolodchikov:1978xm}. Let us consider one of these integrals (see appendix {\bf\ref{Integrals}}, equations (\ref{3p I}), (\ref{3p I alpha}) and (\ref{3p I alpha gamma})): 
\begin{align}
	I &= \iint d^2x \: d^2y \: e^{-i x\cdot \Delta - i y\cdot \tilde{\Delta}} \int \frac{d^2q}{4 \pi^2} \: e^{i q \cdot (x-y)} D(q) \nonumber \\
	&= \left( \dslash{\Delta} + m \right) \left\{ \frac{2\pi}{4 \omega(\Delta)} \left[ \delta \left( \Delta_0 - \omega(\Delta) \right) - \delta \left( \Delta_0 + \omega(\Delta) \right) \right] + \frac{i}{\Delta^2 - m^2} \right\} \: 4 \pi^2 \delta^{(2)}(\mathbf{k} - \mathbf{p}) \label{oneofthem}. 
\end{align}
Strictly speaking, the result in (\ref{oneofthem}) is valid only for the $\Delta_0 - \omega(\Delta) \ne 0$ case. Indeed, this is the case for the standard situation of one pole on a real line.\footnote{This is often formally written as $\frac{1}{x \pm i0} = \mp i \pi \delta(x) + \rm{p.v.} \left( \frac{1}{x} \right)$.} However, a new feature in this model is that one needs to carefully take into account both poles of the propagator in (\ref{oneofthem})\footnote{We remind that although the relativistic propagator is a retarded propagator by the choice of the false vacuum (\ref{AAF pseudovacuum}), one still needs to take into account both poles of the propagator when computing the integrals.}. Let us briefly explain how the computation is done (full details are given in the appendix {\bf \ref{appendix integral I details}}). The contour of the integration over $q_0$ is split into integrations over the segments $(-\infty, -\omega(q)-\epsilon)$, $(-\omega(q)+\epsilon, \omega(q)-\epsilon)$, and $(\omega(q) +\epsilon, +\infty)$, which is what we called the \rm{p.v.} term, and the integrations over the two semi-circles around the two poles at $\pm \omega(q)$, which result in the $\delta$-function terms in (\ref{oneofthem}). A very careful analysis of the \rm{p.v.} term shows that a typical integral to be evaluated has the form: 
\begin{align}
	\int_{0}^{\infty}dx\sin(x \eta)\si(x \epsilon)=\left\{ 
	\begin{array}{rl}
		-\frac{\pi}{2 \eta} ,& \eta^2 > \epsilon^2 \\
		0 ,& \eta^2 < \epsilon^2 
	\end{array}
	\right., \label{typical} 
\end{align}
where $\eta \equiv \Delta_0 - \omega(\Delta)$, and $\si(x)$ is the sine integral (see, for example, (6.252) of \cite{Gradryzhik:1994}). The result in (\ref{oneofthem}) is valid for $\eta \ne 0$, and, therefore, $\eta^2 > \epsilon^2$, since we should consider the limit $\epsilon \rightarrow 0$. Let us note that the $\eta=0$ point corresponds exactly to the integrability condition, namely, to the condition that the set of initial momenta is equal to the set of final momenta. In other words, the result in (\ref{oneofthem}) is valid for the set of momenta which are not at the integrability point. Let us now turn to the case for which $\eta=0$. This corresponds to the integrability point, and the integral (\ref{typical}) is equal to zero. Therefore, one in principle will obtain a different result from the one in (\ref{oneofthem}). This is somewhat puzzling, as one would expect continuity of the scattering amplitude in the external momenta, and this issue should not have come up. It is clear, however, that even though each separate term, as the one above, should not be in principle continuous, the continuity in the external momenta should be restored when the contribution of all diagrams in each order in $\hbar$ is taken into account. The artificial singularity, that arises here in the point $\eta=0$, appears because we split ``by hand'' the scattering amplitude into several terms corresponding to the integrals (\ref{3p I}), (\ref{3p I alpha}) and (\ref{3p I alpha gamma}). 

This special point  makes the analysis technically much more complicated, due to the enormous number of permutations. Indeed, by considering one particular integrability condition, corresponding to one fixed choice of the initial and final set of momenta, one needs to explicitly write down all the terms in (\ref{3p 1-loop 1-contraction off-shell}), consider separately the subset corresponding to $\eta=0$, and the subset for which $\eta \ne 0$. This is quite difficult to deal with, and instead we exclude this special case by the following argument. To be more precise in our analysis, we should, strictly speaking, consider localized wave-packet distributions, corresponding to the scattering particles. In this case the special point $\eta=0$ is never reached, and in fact it should not even be taken into account. We then obtain the result in (\ref{oneofthem}), which is used below to show the factorization of the $S$-matrix. Let us note that the formal proof of $S$-matrix factorizability for the quantum integrable systems requires consideration of such localized wave-packet distributions (see for details \cite{Parke:1980ki,Dorey:1996gd}). Thus, our consideration is in complete agreement with the formal proof of the $S$-matrix factorization, and, therefore, we will use the result in (\ref{oneofthem}), as well as similar results for the integrals in (\ref{3p I}), (\ref{3p I alpha}) and (\ref{3p I alpha gamma}) in appendix {\bf\ref{Integrals}}. 

Finally, we note that this difficulty does not arise in the calculation for simpler models, such as, the Landau-Lifshitz model \cite{Melikyan:2008cy}. This is because the propagator of the Landau-Lifshitz model has only one pole, and the special case $\eta=0$ does not contribute, in other words, the \rm{p.v.} is simply equal to zero. In contrast, in the AAF model there are two poles, and the $\eta=0$ case produces a non-zero \rm{p.v.} contribution. However, as we have explained above by utilizing the localized wave-packet distributions, the \rm{p.v.} should be a continuous function of the external momenta, and $\eta=0$ case plays no role in further analysis. 

\subsection{S-matrix factorization} 
\label{Sfactor}
In this section, we prove $S$-matrix factorization at first non-trivial order by showing that the second term in (\ref{3p tree-level on-shell}) and (\ref{3p 1-loop}) cancel each other. The idea is to rewrite them in terms of rapidities, so that it is possible to compute the spinorial products without resorting to an identity, such as (\ref{on-shell relation for delta}), and then work out the antisymmetrizations. Consider the rapidities: 
\begin{equation*}
	p_0^i = m \cosh \theta_i, \: p_1^i = m \sinh \theta_i, \: k_0^i = m \cosh \eta_i, \: k_1^i = m \sinh \eta_i, \: i=1,2,3. 
\end{equation*}

The non-integrable contribution at tree-level is easily evaluated: 
\begin{align}
	\label{3p tree-level rapidities} -\frac{ig_3}{16m} (3!)^2 \mathcal{A}_{p,k} \left[ k^3 \times p^3 \: \bar{U}^k_{123} \: \mathbb{Q} \: U^p_{123} \right] = -i g_3 m \cosh \left( \sum_{i=1}^3 \frac{ \eta_i - \theta_i}{2} \right) F\left( \boldsymbol{\eta}, \boldsymbol{\theta} \right), 
\end{align}
where we introduced the rapidity-dependent function: 
\begin{equation}
	\label{rapidity F} F\left( \boldsymbol{\eta}, \boldsymbol{\theta} \right) = \frac{\sinh \left( \frac{\eta_1 - \eta_2}{2} \right) \sinh \left( \frac{\eta_1 - \eta_3}{2} \right) \sinh \left( \frac{\eta_2 - \eta_3}{2} \right) \sinh \left( \frac{\theta_1 - \theta_2}{2} \right) \sinh \left( \frac{\theta_1 - \theta_3}{2} \right) \sinh \left( \frac{\theta_2 - \theta_3}{2} \right) }{\sqrt{\cosh \eta_1 \cosh \eta_2 \cosh \eta_3 \cosh \theta_1 \cosh \theta_2 \cosh \theta_3 } }. 
\end{equation}

The next step is to reduce the second term from (\ref{3p 1-loop}) to the opposite of the tree-level contribution (\ref{3p tree-level rapidities}), though in this case the calculations are considerably more involved. First, we recast the argument of the antisymmetrizator in terms of rapidities: 
\begin{align}
	\frac{\left[p^2\times p^1+\Delta\times k^1\right] \left[k^3 \times k^2 + \Delta \times p^3\right]}{\Delta^2 -m^2} \: \bar{U}^k_{123}\mathbb{M}(\Delta)U^p_{213} = -\frac{8m^3 G\left( \boldsymbol{\eta}, \boldsymbol{\theta} \right)} {\sqrt{ \prod_{i=1}^3 \cosh \eta_i \cosh \theta_i } }, 
\end{align}
where 
\begin{align}
	G\left( \boldsymbol{\eta}, \boldsymbol{\theta} \right) = \cosh \left( \frac{\eta_2 - \eta_3}{2} \right) \cosh \left( \frac{\theta_1 - \theta_2}{2} \right) \cosh \left( \frac{\eta_2 - \theta_3}{2} \right) \coth \left( \frac{\eta_1 - \theta_1}{2} \right) \cdot \nonumber \\
	\cdot \sinh \left( \frac{\eta_1 - \theta_2}{2} \right) \sinh \left( \frac{\eta_3 - \theta_3}{2} \right) \sinh \left( \frac{\eta_2 - \eta_3 - \theta_2 + \theta_3}{2} \right). 
\end{align}
Then, instead of directly antisymmetrizing it, it is more profitable if we rewrite $G\left( \boldsymbol{\eta}, \boldsymbol{\theta} \right)$ so as to minimize the number of hyperbolic functions depending only on the difference of one $\eta$ and one $\theta$, obtaining, thus the following decomposition: 
\begin{equation}
	\label{G decomposition} G\left( \boldsymbol{\eta}, \boldsymbol{\theta} \right) = \frac{1}{4} \sum_{i =1}^8 A_i\left( \boldsymbol{\eta}, \boldsymbol{\theta} \right), 
\end{equation}
with the factors $A_i\left( \boldsymbol{\eta}, \boldsymbol{\theta} \right)$ defined in appendix {\bf\ref{Factorizable Tree-Level Details}} by (\ref{1-loop pv A1} - \ref{1-loop pv A8}). It is clear then that we can apply the antisymmetrizator at each $A_i\left( \boldsymbol{\eta}, \boldsymbol{\theta} \right)$ independently. Remarkably, 
\begin{equation}
	\mathcal{A}_{\theta, \eta} \left[ A_i\left( \boldsymbol{\eta}, \boldsymbol{\theta} \right) \right] = 0, \quad \textrm{for} \: i=2,4,5,7 \quad \textrm{and} \quad \mathcal{A}_{\theta, \eta} \left[ A_6\left( \boldsymbol{\eta}, \boldsymbol{\theta} \right) \right] = \mathcal{A}_{\theta, \eta} \left[ A_8\left( \boldsymbol{\eta}, \boldsymbol{\theta} \right) \right]. 
\end{equation}
We give the long explicit expressions for the non-zero terms in the appendix {\bf \ref{Factorizable Tree-Level Details}}. The expression for $G\left( \boldsymbol{\eta}, \boldsymbol{\theta} \right)$ is considerably simplified upon antisymmetrization 
\begin{equation}
	\label{G antisymmetrized} \mathcal{A}_{\theta, \eta} \Big[ G\left( \boldsymbol{\eta}, \boldsymbol{\theta} \right) \Big] = \frac{1}{4} \mathcal{A}_{\theta, \eta} \Big[ A_1\left( \boldsymbol{\eta}, \boldsymbol{\theta} \right) + A_3\left( \boldsymbol{\eta}, \boldsymbol{\theta} \right) + 2 A_6\left( \boldsymbol{\eta}, \boldsymbol{\theta} \right)\Big], 
\end{equation}
but not yet reduced to the opposite of (\ref{3p tree-level rapidities}).

In addition, there is still the condition of overall energy and momentum conservation, which has not been so far imposed on (\ref{G antisymmetrized}). In fact, as we restrict the two-momenta to this submanifold, by implementing the overall delta function $\delta^{(2)}\left( \mathbf{k} - \mathbf{p} \right)$, we have shown that both $\mathcal{A}_{\theta, \eta} \left[ A_3\left( \boldsymbol{\eta}, \boldsymbol{\theta} \right) \right]$ and $\mathcal{A}_{\theta, \eta} \left[ A_6\left( \boldsymbol{\eta}, \boldsymbol{\theta} \right) \right]$ vanish identically. The most straightforward way to do this is to consider light-cone momenta: 
\begin{equation*}
	p_{\pm}^i = e^{\pm \theta} \quad \textrm{and} \quad k_{\pm}^i = e^{\pm \eta} \quad (i=1,2,3), 
\end{equation*}
for which the mass-shell condition is convenient recast as: 
\begin{equation*}
	p_-^i = \frac{1}{p_+^i} \quad \textrm{and} \quad k_-^i = \frac{1}{k_+^i}. 
\end{equation*}
This provides a suitable way not only to write $\mathcal{A}_{\theta, \eta} \left[ A_3\left( \boldsymbol{\eta}, \boldsymbol{\theta} \right) \right]$ and $\mathcal{A}_{\theta, \eta} \left[ A_6\left( \boldsymbol{\eta}, \boldsymbol{\theta} \right) \right]$, as they depend only on hyperbolic functions, but to implement the conservation of energy and momentum without introducing non-linear relations amongst the two-momenta. Then, after very long and tedious calculations one finds: 
\begin{equation}
	\label{A3 = A6 = 0} \mathcal{A}_{\theta, \eta} \left[ A_3\left( \boldsymbol{\eta}, \boldsymbol{\theta} \right) \right] \delta^{(2)} \left( \mathbf{k} - \mathbf{p} \right) = 0 \quad \textrm{and} \quad \mathcal{A}_{\theta, \eta} \left[ A_6\left( \boldsymbol{\eta}, \boldsymbol{\theta} \right) \right] \delta^{(2)} \left( \mathbf{k} - \mathbf{p} \right) = 0. 
\end{equation}
Thus, we obtain: 
\begin{align}
	\left( 3! \right)^2 \mathcal{A}_{\theta, \eta} \Big[ G\left( \boldsymbol{\eta}, \boldsymbol{\theta} \right) \Big] \delta^{(2)} \left( \mathbf{k} - \mathbf{p} \right) &= \frac{1}{4} \left( 3! \right)^2 \mathcal{A}_{\theta, \eta} \Big[ A_1\left( \boldsymbol{\eta}, \boldsymbol{\theta} \right) - A_3\left( \boldsymbol{\eta}, \boldsymbol{\theta} \right) \Big] \delta^{(2)} \left( \mathbf{k} - \mathbf{p} \right) \nonumber \\
	&= 2 \cosh \left( \sum_{i=1}^3 \frac{\eta_i - \theta_i}{2} \right) \sinh \left( \frac{\eta_1 - \eta_2}{2} \right) \sinh \left( \frac{\eta_1 - \eta_3}{2} \right) \cdot \nonumber \\
	&\cdot \sinh \left( \frac{\eta_2 - \eta_3}{2} \right) \sinh \left( \frac{\theta_1 - \theta_2}{2} \right) \sinh \left( \frac{\theta_1 - \theta_3}{2} \right) \cdot \nonumber \\
	&\cdot \sinh \left( \frac{\theta_2 - \theta_3}{2} \right) \delta^{(2)} \left( \mathbf{k} - \mathbf{p} \right). 
\end{align}
And finally, 
\begin{align}
	\label{3p 1-loop rapidities} i \left(-\frac{3!\: ig_2}{4m}\right)^2 \mathcal{A}_{k,p} \left\{ \frac{\left[p^2\times p^1+\Delta\times k^1\right] \left[k^3 \times k^2 + \Delta \times p^3\right]}{\Delta^2 -m^2} \bar{U}^k_{123}\mathbb{M}(\Delta)U^p_{213}\right\} 4\pi^2 \delta^{(2)} \left( \mathbf{k} - \mathbf{p} \right)= \nonumber \\
	= i g_2^2 m \cosh \left( \sum_{i=1}^3 \frac{ \eta_i - \theta_i}{2} \right) F\left( \boldsymbol{\eta}, \boldsymbol{\theta} \right) 4 \pi^2 \delta^{(2)} \left( \mathbf{k} - \mathbf{p} \right). 
\end{align}
Noting also that there should be a factor of $4 \pi \delta^{(2)} \left( \mathbf{k} - \mathbf{p} \right)$ multiplying the tree-level non-integrable term (\ref{3p tree-level rapidities}), we can easily conclude that the term which prevented $S$-matrix factorization at tree level (\ref{3p tree-level rapidities}) is indeed cancelled by the \rm{p.v.} contribution coming from the one-contraction diagrams at one-loop (\ref{3p 1-loop rapidities}), provided the following constraint on the coupling constants holds: 
\begin{equation}
	g_2^2 = g_3. \label{superconstraint}
\end{equation}
This is one of our central results, which provides the necessary condition for the quantum integrability of the AAF model. Note that this condition is in complete agreement with the mass dimensions assigned to the coupling constants $g_2$ and $g_3$, and the classical Lagrangian (\ref{AAF action}), as discussed in section {\bf \ref{AAF QFT}}. Indeed, with (\ref{superconstraint}), one can easily go back from the Lagrangian (\ref{AAF action}) with coupling constants to its dimensionless counterpart (\ref{AAF S mass action}) with the correct coefficients. 

Therefore, if we denote $g=g_2$, the scattering amplitude for the three-particle scattering becomes 
\begin{equation}
	\label{3p factorizable S matrix} \langle \mathbf{k} | \hat{S} | \mathbf{p} \rangle = \left\{ 1 + 2 \sum_{n=1}^2 \left[ \left( -\frac{i g}{4m} \right) \Big( p^2 \times p^1 + p^3 \times p^1 + p^3 \times p^2 \Big) \right]^n \right\} \langle \mathbf{k} | \mathbf{p} \rangle + O(g^3), 
\end{equation}
where we included the one-loop amplitudes $V^{(2)}_i \left( \mathbf{k}, \mathbf{p} \right)$, $i=2,3,4$ in $O(g^3)$, for 
\begin{equation*}
	V^{(2)}_2 \left( \mathbf{k}, \mathbf{p} \right) \sim V^{(2)}_3 \left( \mathbf{k}, \mathbf{p} \right) \sim g_2 g_3 \sim g^3 \quad \textrm{and} \quad V^{(2)}_4 \left( \mathbf{k}, \mathbf{p} \right) \sim g_3^2 \sim g^4. 
\end{equation*}
The first term in (\ref{3p factorizable S matrix}) corresponds to the factorizable\footnote{Compare (\ref{3p factorizable S matrix}) with (\ref{3p S matrix from 2p S matrix}), which we derived only from the knowledge of the two-particle $S$-matrix.} three-particle $S$-matrix at second order in $g$. Consequently, in order to have $S$-matrix factorization even at tree-level, it is mandatory to consider higher loop amplitudes, as they yield the counterterms for lower order non-integrable terms. 

It is not difficult to see that this remarkable scheme of cancellations would have been completely ruined had we not corrected the factor of $\frac{1}{2}$ missed by \cite{Alday:2005jm} and the overall sign difference in comparison with by \cite{Klose:2006dd}, as they were paramount for the intricate fine-tuning between the interaction terms required for $S$-matrix factorization. Rectifying the signs in the action (\ref{AAF action}) had profound consequences in the analysis thereafter, the most notable being that instead of deriving the $S$-matrix proposed by \cite{Klose:2006dd} we obtained its inverse. Remarkably, our two-particle $S$-matrix (\ref{2p S matrix}), written in terms of rapidities: 
\begin{equation}
	\label{2p S matrix rapidities} S(\theta_1,\theta_2) = \frac{1 - \frac{img_2}{4} \sinh\left(\theta_1 - \theta_2 \right)}{1 + \frac{img_2}{4} \sinh\left(\theta_1 - \theta_2 \right)}, 
\end{equation}
is very similar to the $S$-matrix for two-particle scattering of the massive Thirring model: 
\begin{equation}
	\label{Thirring S matrix} S_{Thirring}(\theta, \theta') = \frac{1 - \frac{ig}{2} \tanh\frac{\theta - \theta'}{2}}{1 + \frac{ig}{2} \tanh\frac{\theta - \theta'}{2}}. 
\end{equation}
We can now write down the correct Bethe equations for the AAF model:
\begin{equation}
	\label{Bethe equations} e^{i J m \sinh \theta_i} =\prod_{k \neq i} \frac{1 + \frac{img_2}{4} \sinh\left(\theta_i - \theta_k \right)}{1 - \frac{img_2}{4} \sinh\left(\theta_i - \theta_k \right)}, 
\end{equation}
which is the first step in analyzing the bound and negative energy states, and obtaining the physical $S$-matrix and excitations for both repulsive and attractive cases \cite{Klose:2006dd,Korepin:1979qq,Korepin:1979hg}.

We conclude the paper by making a comparison with the analysis and the results of \cite{Klose:2006dd}. Towards this end, we start from the original AAF action (\ref{AAF action lorentz invariant}) and consider the transformation:\footnote{Here we also neglect the constant term, fix $\kappa = \frac{\sqrt{\lambda}}{2}$ and change the Dirac matrices basis through (\ref{dirac matrices similarity tranformation}).}
\begin{equation}
	\label{f_transform}
	\tau \rightarrow -\tau, \quad \sigma \rightarrow -\sigma, \quad \rho^{\alpha} \rightarrow -\rho^{\alpha},
\end{equation}
Then the action becomes:
\begin{align} 
	\label{parity} S &= \frac{\sqrt{\lambda}}{2 \pi} \int d\tau \: \int_0^{\frac{2 \pi J}{\sqrt{\lambda}}} d\sigma \: \left[ \frac{i}{2} \left(\bar{\chi} \: \gamma^{\alpha} 
	\partial_{\alpha} \chi - 
	\partial_{\alpha} \bar{\chi} \: \gamma^{\alpha} \chi \right) - \bar{\chi} \chi \: - \right. \nonumber \\
	&- \left. \frac{1}{4}\epsilon^{\alpha \beta} \left( \bar{\chi}
	\partial_{\alpha} \chi \; \bar{\chi}\: \gamma^3 
	\partial_{\beta} \chi -
	\partial_{\alpha}\bar{\chi} \chi \; 
	\partial_{\beta} \bar{\chi}\: \gamma^3 \chi \right) - \frac{1}{16} \epsilon^{\alpha \beta} \left(\bar{\chi}\chi\right)^2 
	\partial_{\alpha}\bar{\chi}\:\gamma^3
	\partial_{\beta}\chi \right], 
\end{align}
where we have denoted $\chi(\tau,\sigma) \equiv \psi(-\tau,-\sigma)$. This action differs from our action (\ref{AAF S action symmetric}) by the sign of the quartic term, and this is essentially the action considered by \cite{Klose:2006dd}, written in terms of the field $\chi(\tau,\sigma)$. Thus, the two actions are different already on the classical level. Careful analysis shows that the $S$-matrix of \cite{Klose:2006dd} corresponds to the scattering of the $\chi$-particles, rather than the original $\psi$-particles. It is easy to see from the mode expansions for  both fields, that going from $\chi(\tau,\sigma)$ to $\psi(\tau,\sigma)$ corresponds to interchanging the particle and anti-particle operators $a^{\dagger} \longleftrightarrow b^{\dagger}$, as the transformation (\ref{f_transform}) involves time inversion. Hence, it is not surprising that our $S$-matrix (\ref{2p S matrix rapidities}) is the inverted result of \cite{Klose:2006dd}, and to make a connection between the two results we must change the coupling constant $g_2 \rightarrow -g_2$. Indeed, our general result (\ref{2p S matrix rapidities}) explicitly shows this. 

It is important to emphasize, that our quantum integrability condition (\ref{superconstraint}) is invariant under the transformation $g_2 \rightarrow -g_2$, and consequently, both models (\ref{AAF action}) and (\ref{parity}) are integrable. Finally, we stress that even though this action arises from string theory, our generalized  model with two independent coupling constants is an interesting case to investigate on its own, where both repulsive and attractive cases should be considered separately \cite{Korepin:1979qq,Korepin:1979hg}.

\section{Conclusion}\label{Conclusion}

In this paper we have considered the quantum integrability of the AAF model, and showed the S-matrix factorizability in the first non-trivial order. As we explain in the main text, the AAF model requires introducing two dimensional coupling constants, and one of our main results is a necessary relation between these coupling constants in order to guarantee the quantum integrability. With this quantum constraint we were able to reveal and correct several missed factors in the previous works, as well as to derive the correct S-matrix. The latter is the inverse of the one found in \cite{Klose:2006dd}. This also changes the analysis of bound and negative energy states.  In the process, the mechanism behind the cancellations of non-integrable parts is understood at the perturbative level. 

As discussed in the introduction, the AAF model is a very interesting fermionic integrable model, which is considerably more complex than its simpler fermionic Thirring model counterpart. While for the latter we have a number of techniques to understand its quantization, for the AAF model this is not the case due to its complexity. Even though the quantum inverse scattering method is a relatively straightforward procedure, in the AAF model the main difficulty lies in the type of the interaction Hamiltonian, namely, in the singular behavior of the quantum mechanical Hamiltonian and all other conserved charges. Let us note, that in the Landau-Lifshitz model one encounters exactly the same type of singularity, and the development of the quantum inverse scattering method required  considerable effort and careful analysis of operator products and their regularization, as well as the construction of the correct Hilbert space \cite{Sklyanin:1988,Melikyan:2008ab,Melikyan:2010bi,Melikyan:2010fr}. It has also been shown that the self-adjointness of the operators is essentially equivalent to the S-matrix factorization. It is desirable to perform a similar analysis for the AAF model, and we plan to do it in the future. The lattice version of the AAF model should be also understood in the process. It would be also interesting to compare the two extended Hilbert spaces for the two models. Indeed, we do not expect that the constructions of the self-adjoint operators and corresponding extensions should coincide, as they carry a different number of degrees of freedom. Moreover, one should understand whether the two extensions can be accommodated in some larger space to fit both fermionic and bosonic degrees of freedom. This will be a very important step in understanding the correct Hilbert space for the entire string on $AdS_5 \times S^5$, as we have emphasized in \cite{Melikyan:2008ab,Melikyan:2010bi,Melikyan:2010fr}.

While the AAF model has some similarity to the Landau-Lifshitz model, there are a few distinctions that make the AAF model a more intriguing theory. In particular, the Poisson brackets structure in the AAF model is highly non-linear, due to the presence of the time derivatives in the forth and sixth order of the interaction vertices, extending up to the eighth order in fermionic fields, which makes it quite difficult to develop the standard quantum inverse scattering method. While there are examples of such models, for instance the \emph{anisotropic} Landau-Lifshitz model where the standard commutation relations between the fields should be modified in the quantum theory, resulting in non-linear \emph{Sklyanin algebra}, it is hard to deal with such theories. Besides that, there is a deep relation between the algebraic structure and regularization of the singular Yang-Baxter relations. In addition, the non-linearity in the commutation relations in the AAF model already appears in the classical theory. This is somewhat unusual, and may lead to the loss of some non-perturbative effects in the perturbative analysis. For other known models this does not happen, and the S-matrix perturbative calculations have produced consistent results. However, strictly speaking, the perturbative S-matrix calculations are not reliable, and the full picture can be understood only within the framework of the quantum inverse scattering method. Therefore, developing the latter, together with the careful analysis of bound and negative states, as well as construction of the physical S-matrix and excitations, should be the main focus in the future investigations. 

\section*{Acknowledgements}
We thank S. Frolov and T. Klose for useful comments. 

A.P. acknowledges partial support of CNPq under
grant no.308911/2009-1. The work of V.O.R. is supported by CNPq grant 304116/2010-6 and FAPESP grant 2008/05343-5. The work of G.W. is supported by the FAPESP grant No. 06/02939-9.

\section*{Appendices} \addcontentsline{toc}{section}{Appendices}
\appendix 
\section{Two-dimensional Dirac equation}\label{Appendix Dirac}

Consider the two-dimensional Dirac equation: 
\begin{equation}
	\label{apendix dirac 2d equation} \left( i \gamma^{\mu} 
	\partial_{\mu} -m \right)\psi(x) = 0, 
\end{equation}
with the Dirac matrices $\gamma^{\mu}$, $\mu =0,1$, belonging to the $SO(1,1)$ Clifford algebra: 
\begin{equation}
	\label{clifford SO(1,1) algebra} \gamma^{\mu}\gamma^{\nu} + \gamma^{\nu}\gamma^{\mu} = \eta^{\mu \nu} \mathbb{1}_2\:, 
\end{equation}
where the two-dimensional Minkowsky metric is $\eta = \diag(1,-1)$ and the symbol $\mathbb{1}_2$ stands for the $2 \times 2$ unit matrix. In the main text, we consider the following faithful representations of (\ref{clifford SO(1,1) algebra}): 
\begin{equation}
	\label{dirac matrices rho} \rho^0 = \left( 
	\begin{array}{cc}
		-1 & 0 \\
		0 & 1 
	\end{array}
	\right), \quad \rho^1 = \left( 
	\begin{array}{cc}
		0 & i \\
		i & 0 
	\end{array}
	\right), \quad \rho^5 = \rho^0 \rho^1,
\end{equation}
and 
\begin{equation}
	\label{dirac matrices gamma} \gamma^0 = \left( 
	\begin{array}{cc}
		0 & 1 \\
		1 & 0 
	\end{array}
	\right), \quad \gamma^1 = \left( 
	\begin{array}{cc}
		0 & -1 \\
		1 & 0 
	\end{array}
	\right), \quad \gamma^3 = \gamma^0 \gamma^1,
\end{equation}
which are related by the similarity transformation bellow: 
\begin{equation}
	\label{dirac matrices similarity tranformation} \gamma^{\mu} = M \rho^{\mu} M^{-1}, \quad M = \frac{1}{\sqrt{2}} \left( 
	\begin{array}{cc}
		1 & -i \\
		-1 & -i 
	\end{array}
	\right). 
\end{equation}

\subsection{Plane-wave solutions}

Substituting the plane-wave solution: 
\begin{equation}
	\psi_a(x) = e^{-i p \cdot x} u_a(p), \quad \mathrm{with} \quad p \cdot x = \eta^{\mu \nu} p_{\mu} x_{\nu}. 
\end{equation}
into (\ref{apendix dirac 2d equation}) leads to the equation of motion for the spinor: 
\begin{equation}
	\label{appendix dirac 2d spinor positive eq} \left(\dslash{p} - m\right) u_a(p) = 0, 
\end{equation}
which has non-trivial solutions, if and only if, 
\begin{equation*}
	\det\left( \dslash{p} - m\right) = 0 \Rightarrow m^2 - p^2 = 0, 
\end{equation*}
i.e., if and only if, the momentum $p$ is on mass-shell, solving this condition for the energy, we get: 
\begin{equation}
	\label{appendix dirac 2d mass shell} p_0 = \sqrt{p_1^2 + m^2} =: \omega(p). 
\end{equation}

The normalized solutions of (\ref{appendix dirac 2d spinor positive eq}), with positive and negative energy, are: 
\begin{equation}
	\label{appendix dirac 2d spinors} u(p) = \left( 
	\begin{array}{c}
		\sqrt{\frac{\omega(p)-p_1}{2\omega(p)}} \\
		\sqrt{\frac{\omega(p)+p_1}{2\omega(p)}} 
	\end{array}
	\right), \quad v(p) = \left( 
	\begin{array}{c}
		\sqrt{\frac{\omega(p)+p_1}{2\omega(p)}} \\
		-\sqrt{\frac{\omega(p)-p_1}{2\omega(p)}} 
	\end{array}
	\right).
\end{equation}
Therefore, the plane-wave solutions for the Dirac equation in two dimensions are: 
\begin{equation}
	\label{appendix dirac 2d plane wave solutions} 
	\begin{array}{ll}
		\psi_+(x) = e^{-ip\cdot x}\:u(p),& \mathrm{with} \quad \left(\dslash{p}-m\right)u(p) = 0, \\
		\psi_-(x) = e^{ip\cdot x}\:v(-p),& \mathrm{with} \quad \left(\dslash{p}+m\right)v(-p) = 0. 
	\end{array}
\end{equation}

\subsection{Completeness and orthonormality relations}

The solutions $u(p)$ and $v(p)$ (\ref{appendix dirac 2d spinors}) satisfy the  orthonormality relations: 
\begin{equation}
	\label{appendix dirac 2d orthonormaly relations} 
	\begin{array}{ll}
		u^{\dagger}(p)u(p) = 1 ,& v^{\dagger}(p)v(p) = 1, \\
		u^{\dagger}(p)v(p) = 0 ,& v^{\dagger}(p)u(p) = 0, 
	\end{array}
\end{equation}
as well as, 
\begin{equation}
	\label{appendix dirac 2d spinorial relations} 
	\begin{array}{ll}
          \bar{u}(p)u(p) = \frac{m}{\omega(p)} ,& \bar{v}(p)v(p) = -\frac{m}{\omega(p)}, \\
          \bar{u}(p)v(p) = \frac{p_1}{\omega(p)} ,& \bar{v}(p)u(p) = \frac{p_1}{\omega(p)}, 
	\end{array}
\end{equation}
where the conjugate spinors are defined in the usual way: $\bar{u}= u^{\dagger}\gamma^0$ and $\bar{v}= v^{\dagger}\gamma^0$. 

Finally, the completeness relations for the spinors $u(p)$ e $v(p)$ are: 
\begin{equation}
	\label{appendix dirac 2d completeness relations} u(p)\bar{u}(p) = \frac{\dslash{p} +m}{2 \omega(p)}, \quad v(-p)\bar{v}(-p) = \frac{\dslash{p} -m}{2 \omega(p)}. 
\end{equation}

\section{Table of useful integrals}\label{Integrals} Let $D(q)$ be the propagator in momentum space: 
\begin{equation}
	D(q) = \frac{i \left(\dslash{q} + m \right)}{q^2 - m^2 + 2 i \epsilon q_0}. 
\end{equation}
The relevant momentum space integrals for two-particle scattering are: 
\begin{align}
	I_0(p^1,p^2) &= \int \frac{d^2q}{4 \pi^2} \: D(q) \otimes D(p^1+p^2-q) \nonumber \\
	&= \frac{p^1_0 + p^2_0}{4 |p^1_1 - p^2_1| (p^1 + p^2)^2} \left[ \left( \dslash{p}^1 +m \right)\otimes \left( \dslash{p}^2 +m \right) + \left( \dslash{p}^2 +m \right) \otimes \left( \dslash{p}^1 +m \right) \right] +D_0, \label{I_0}\\
	I_1(p^1,p^2) &= \int \frac{d^2q}{4 \pi^2} \: \left[\left(p^1+p^2-q\right) \times q\right] D(q) \otimes D(p^1+p^2-q) \nonumber \\
	&= \frac{p^2 \times p^1 (p^1_0 + p^2_0)}{4 |p^1_1 - p^2_1| (p^1 + p^2)^2} \left[ \left( \dslash{p}^1 +m \right)\otimes \left( \dslash{p}^2 +m \right) - \left( \dslash{p}^2 +m \right) \otimes \left( \dslash{p}^1 +m \right) \right] +D_1, \label{I_1} \\
	I_2(p^1,p^2) &= \int \frac{d^2q}{4 \pi^2} \: \left[\left(p^1+p^2-q\right) \times q\right]^2 D(q) \otimes D(p^1+p^2-q) \nonumber \\
	&= \frac{ \left(p^2 \times p^1\right)^2 (p^1_0 + p^2_0)}{4 |p^1_1 - p^2_1| (p^1 + p^2)^2} \left[ \left( \dslash{p}^1 +m \right)\otimes \left( \dslash{p}^2 +m \right) + \left( \dslash{p}^2 +m \right) \otimes \left( \dslash{p}^1 +m \right) \right] +D_2, \label{I_2} 
\end{align}
where $D_i$, $i =0,1,2$ stand for their respective divergent parts.

For three-particle scattering the following integrals are also needed: 
\begin{align}
	I &= \iint d^2x \: d^2y \: e^{-i x\cdot \Delta - i y\cdot \tilde{\Delta}} \int \frac{d^2q}{4 \pi^2} \: e^{i q \cdot (x-y)} D(q) \nonumber \\
	&= \left( \dslash{\Delta} + m \right) \left\{ \frac{2\pi}{4 \omega(\Delta)} \left[ \delta \left( \Delta_0 - \omega(\Delta) \right) - \delta \left( \Delta_0 + \omega(\Delta) \right) \right] + \frac{i}{\Delta^2 - m^2} \right\} \: 4 \pi^2 \delta^{(2)}(\mathbf{k} - \mathbf{p}) \label{3p I}, \\
	I_{\alpha} &= \iint d^2x \: d^2y \: e^{-i x\cdot \Delta - i y\cdot \tilde{\Delta}} \int \frac{d^2q}{4 \pi^2} \: e^{i q \cdot (x-y)} q_{\alpha} D(q) \nonumber \\
	&= \Delta_{\alpha} \left( \dslash{\Delta} + m \right) \left\{ \frac{2\pi}{4 \omega(\Delta)} \left[ \delta \left( \Delta_0 - \omega(\Delta) \right) - \delta \left( \Delta_0 + \omega(\Delta) \right) \right] + \frac{i}{\Delta^2 - m^2} \right\} \: 4 \pi^2 \delta^{(2)}(\mathbf{k} - \mathbf{p}) \label{3p I alpha}, \\
	I_{\alpha \gamma} &= \iint d^2x \: d^2y \: e^{-i x\cdot \Delta - i y\cdot \tilde{\Delta}} \int \frac{d^2q}{4 \pi^2} \: e^{i q \cdot (x-y)} q_{\alpha} q_{\gamma} D(q) \nonumber \\
	&= \Delta_{\alpha} \Delta_{\gamma}\left( \dslash{\Delta} + m \right) \left\{ \frac{2\pi}{4 \omega(\Delta)} \left[ \delta \left( \Delta_0 - \omega(\Delta) \right) - \delta \left( \Delta_0 + \omega(\Delta) \right) \right] + \frac{i}{\Delta^2 - m^2} \right\} \: 4 \pi^2 \delta^{(2)}(\mathbf{k} - \mathbf{p}) \label{3p I alpha gamma}, 
\end{align}
where $\Delta = p^1 + p^2 - k^1$ and $\tilde{\Delta} = p^3 - k^2 -k^3$.

\subsection{Integrals for two-particle scattering: computational details}
In this small section, we quickly outline the main steps involved in the evaluation of (\ref{I_0}) and comment on some of the nuances of the result. We note that the presence of factors such as $\left(p^1+p^2-q\right) \times q$ in the other two integrals, i.e., (\ref{I_1}) and (\ref{I_2}), do not introduce any serious technical complication, despite increasing the superficial degree of divergence. 

Before going into the details of this calculation, we would like to stress that despite the apparent symmetry with respect to the momenta
$p^1$ and $p^2$ in (\ref{appendix AAF integrals 2p 2}), which manifests in the dependence of the integral
only on the sum $p^1+p^2$, this is not the case. Indeed, as we already
discussed in the beginning of the section \ref{2 Particle Scattering}, one must choose a
particular ordering for the incoming momenta $p^1_1$ and $p^2_1$.\footnote{In this paper we choose $p^1_1 > p^2_1.$} Therefore,
after integration over $q_0$, the position of the poles will depend on
this ordering, and, as a consequence, the final answer may depend not only on the sum $p^1+p^2$. In fact, this is the case, as the explicit
calculations below show.

\begin{figure}
  \centering 
  \includegraphics[scale=0.65]{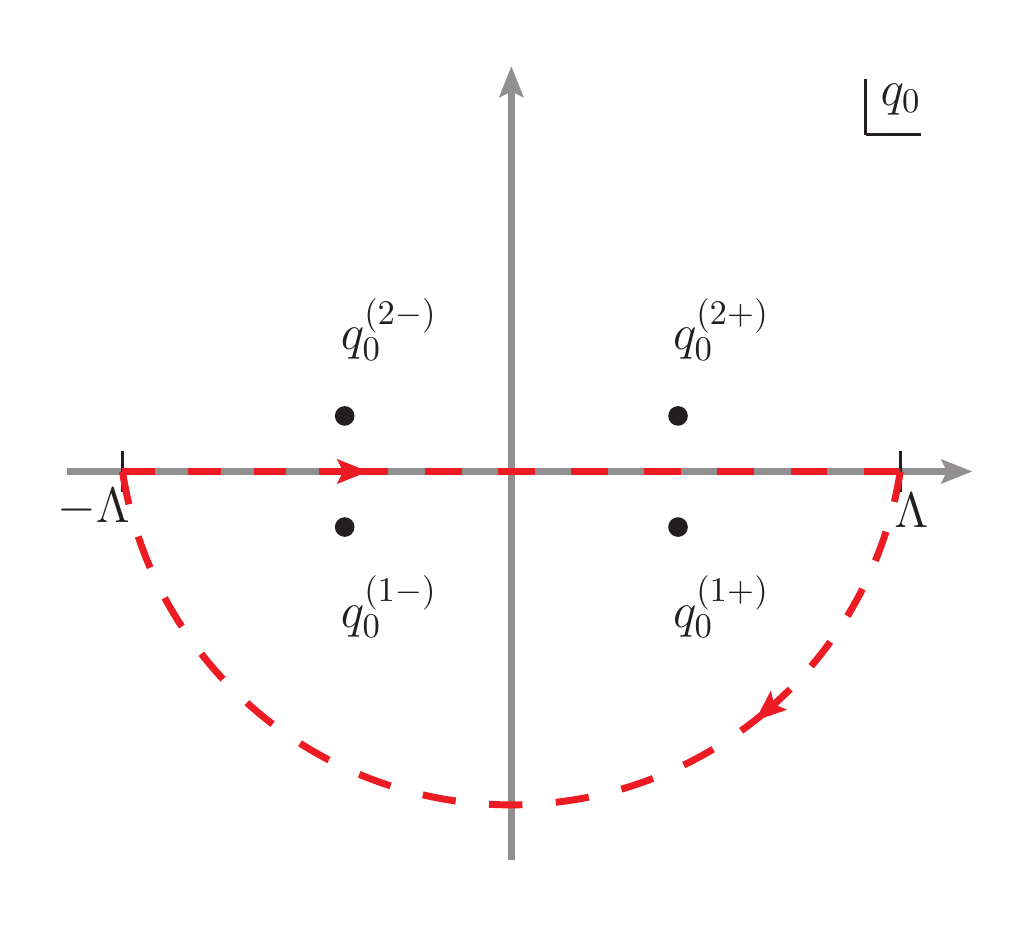} \caption{Pole prescription and contour of integration for computing the integral over $q_0$ in (\ref{appendix AAF integrals 2p 2}).}\label{intpath2p1}   
\end{figure}

We can easily perform the integration over $q_0$ of the integral
\begin{align}\label{appendix AAF integrals 2p 2}
  I_0(p^1,p^2) = \int \frac{d^2q}{4 \pi^2}\: \frac{i \left( \dslash{q} +
    m\right)}{q^2 - m^2 + 2 i \epsilon q_0} \otimes \frac{i \left(
    \dslash{p}^1 + \dslash{p}^2 - \dslash{q}+
    m\right)}{\left(p^1+p^2-q\right)^2 - m^2 + 2 i \epsilon
  \left(p^1_0 + p^2_0 -q_0\right)},
\end{align}
by closing the integration contour in the lower complex half-plane, so that it encloses two of the four simple poles of (\ref{appendix AAF integrals 2p 2}), as depicted in figure \ref{intpath2p1}. The remaining integration over $q_1$ can then be reduced to the following form:
\begin{align}\label{appendix AAF integrals 2p 17}
  I_0(p^1,p^2) = \frac{1}{4 i \pi \left(
      p^1+p^2 \right)^2} \int dq_1 \: \frac{f(q_1)}{\left( q_1 - p^1_1 - i \eta \right) \left(
      q_1 - p^2_1 + i {\eta} \right) },
\end{align}
where $f(q_1)$ is a polynomial in $q_1$ of degree two with some coefficients, but still symmetric in the external momenta. It is important to notice that $\eta$ is a function of $\epsilon$ and the external momenta, which has the following form:
\begin{equation*}
  \eta = \frac{4\epsilon p^1_0p^2_0 \left( p^1_0 + p^2_0
    \right)}{\left( p^1_1 - p^2_1 \right) \left( p^1 + p^2 \right)^2}.
\end{equation*}
From this expression it is clear that in order to perform the remaining integration over $q_1$ one must choose a concrete ordering of the incoming momenta, which fixes the positions of the poles, and ensures that $\eta$ is a well-behaved function. This in turn breaks the symmetry between $p^1$ and $p^2$, as it is clear from the denominator of (\ref{appendix AAF integrals 2p 17}). 

Noting that
\begin{align}\label{appendix AAF integrals 2p 18}
  \int \frac{dq\:\left(a q^2 + b q + c\right)}{\left( q - p^1_1 - i\eta \right) \left(
      q - p^2_1 + i\eta \right)} &= \left[ a\left( P_-^2 +
      P_+^2 \right) + b P_+ + c \right]\int \frac{dx}{\left( x - P_-
      - i \eta \right) \left( x + P_- + i \eta \right)} \: +
  \nonumber \\  
  &+ a \int dx, 
\end{align}
where we wrote $f(q) = aq^2 + bq +c$ and introduced
\begin{equation*}
  P_+ := \frac{p^1_1 + p^2_1}{2} \quad \text{and} \quad  P_- :=
  \frac{p^1_1 - p^2_1}{2},
\end{equation*}
we can finally use
\begin{equation}\label{useful integral}
  \int \frac{dx}{\left( x - P_- - i \eta \right) \left( x + P_-
      + i \eta \right)} = \frac{i\pi}{\left| P_- \right|}
\end{equation}
to obtain the result (\ref{I_0}), with
\begin{equation}
  D_0 = 2 a \lim_{\Lambda \to \infty} \Lambda \: \text{ where } \: a = -\left( p^1_0 + p^2_0 \right) \left( \gamma^0 \otimes \gamma^0 + \gamma^1 \otimes \gamma^1 \right) + \left( p_1^1 + p^2_1 \right) \left( \gamma^0 \otimes \gamma^1 + \gamma^1 \otimes \gamma^0 \right).
\end{equation}

\subsection{Integrals for three-particle scattering: computational details}\label{appendix integral I details}

The computation of the integrals (\ref{3p I} - \ref{3p I alpha gamma}) is considerably more involved and exhibits some interesting features, which are essential for demonstrating the $S$-matrix factorization property. For the reader's convenience, we outline here the main steps for evaluating (\ref{3p I}). We note that the other two integrals, namely (\ref{3p I alpha}) and (\ref{3p I alpha gamma}), can be computed by exactly the same method, and the extra factors of momentum introduce no serious technical complication.

Introducing light-cone-like coordinates: $x = x_+ + x_-$ and $y = x_+ - x_-$, (\ref{3p I}) becomes 
\begin{equation}
	I = 4 \iint d^2x_+ \: d^2x_- \: e^{-i x_+ \cdot \left( \Delta + \tilde{\Delta} \right)} e^{-i x_- \cdot \left( \Delta - \tilde{\Delta} \right)} \int \frac{d^2q}{4 \pi^2} \: e^{2i q\cdot x_-} \frac{i \left( \dslash{q} +m \right)}{q^2 - m^2 + 2 i \epsilon q_0}. 
\end{equation}
We can then integrate over $x_+$, to obtain the overall energy-momentum conservation delta function $\delta^{(2)} \left(\Delta + \tilde{\Delta} \right) = \delta^{(2)} \left( \mathbf{k} - \mathbf{p} \right)$. Rescaling, $x_- \to - \frac{1}{2} x$ and using the exact decomposition for the propagator: 
\begin{equation}
	\frac{1}{q^2 - m^2 + 2 i \epsilon q_0} = \frac{1}{2 \omega(q)} \left[ \frac{1}{q_0 - \omega(q) + i\epsilon} - \frac{1}{q_0 + \omega(q) + i\epsilon} \right], 
\end{equation}
we obtain: 
\begin{equation}
	\label{appendix integral I 1} I = \int d^2x \: e^{ix \cdot \Delta} \int \frac{d^2q}{4 \pi^2} \: \frac{i \left( \dslash{q} + m\right) e^{-i x\cdot q}}{2 \omega(q)} \left[ \frac{1}{q_0 - \omega(q) + i\epsilon} - \frac{1}{q_0 + \omega(q) + i\epsilon} \right] 4 \pi^2 \delta^{(2)}\left( \mathbf{k} -\mathbf{p} \right). 
\end{equation}
\begin{figure}
	\centering 
	\includegraphics[scale=0.65]{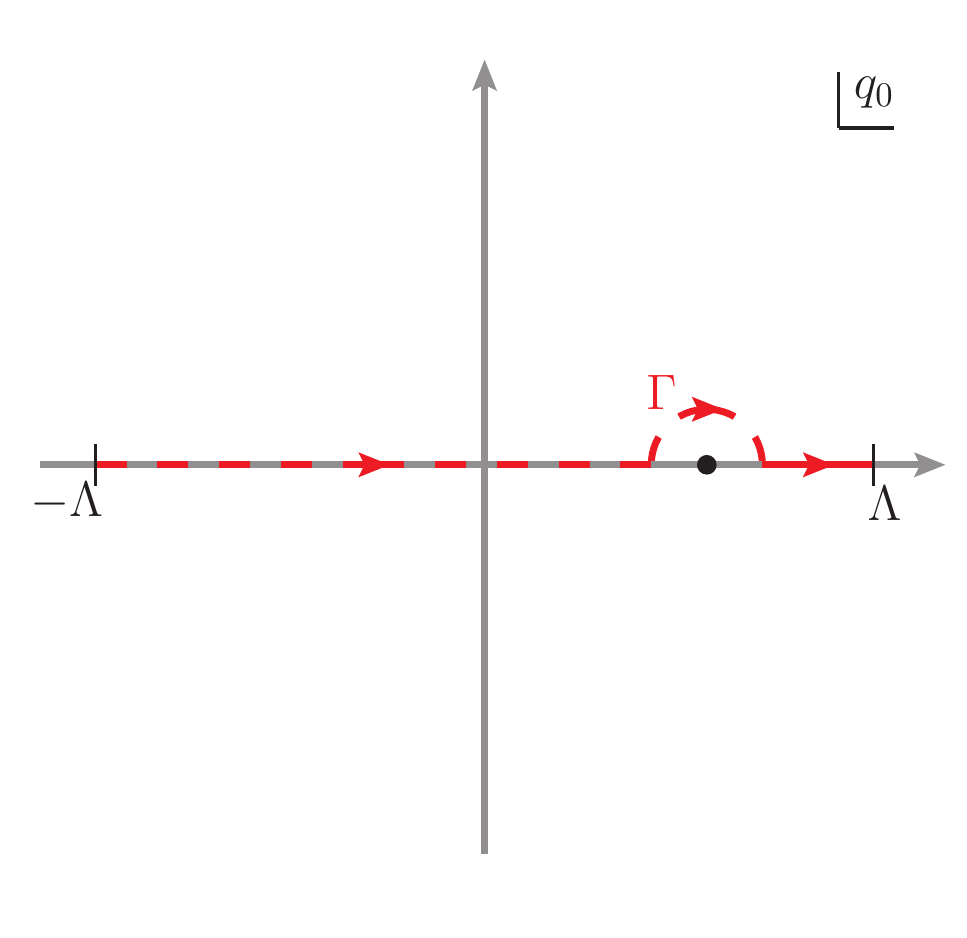} \caption{Integration path $\Gamma$ for the first term in (\ref{appendix integral I 1}).}\label{intpath} 
\end{figure}

For computing the integral over $q_0$ of the first term in (\ref{appendix integral I 1}), we introduce the integration path: 
\begin{equation}
	\Gamma = \left\{ 
	\begin{array}{ll}
		q_0\:,& q_0 \in \Big[ -\Lambda, \omega(q) - \epsilon \Big) \bigcup \Big( \omega(q) + \epsilon, \Lambda \Big] \\
		q_0 -\omega(q) = \epsilon e^{- i \theta}\:,& \theta \in \left[ 0, \pi \right] \quad \left(C_{\epsilon}\right) 
	\end{array}
	\right., 
\end{equation}
depicted in figure \ref{intpath}. Then, 
\begin{align}
	\int \frac{dq_0}{2 \pi} \: \frac{\left( \dslash{q} +m \right) e^{-i x^0 q_0}}{q_0 - \omega(q) + i \epsilon} &= \lim_{\substack{\Lambda \to \infty \\
	\epsilon \to 0}} \left\{ \left[ \int_{-\Lambda}^{\omega(q) - \epsilon} + \int^{\Lambda}_{\omega(q) + \epsilon} \right] \frac{dq_0}{2 \pi} \: \frac{\left( \dslash{q} +m \right) e^{-i x^0 q_0}}{q_0 - \omega(q)} \right. + & \nonumber \\
	&+ \left. \frac{1}{2 \pi} \left[ \left( \dslash{q} +m \right) e^{-i x^0 q_0}\right]\Big|_{q_0 = \omega(q)} \int_{C_{\epsilon}} \frac{dq_0}{q_0 - \omega(q)} \right\} \nonumber \\
	&= \fint \frac{dq_0}{2 \pi} \: \frac{\left( \dslash{q} +m \right) e^{-i x^0 q_0}}{q_0 - \omega(q)} - \frac{i}{2} \left[ \left( \dslash{q} +m \right) e^{-i x^0 q_0} \right]\Big|_{q_0 = \omega(q)}, 
\end{align}
which is nothing but the sum of a principal value term with a delta term that gives rise to the decomposition (\ref{delta + pv split}). The second term of (\ref{appendix integral I 1}) can be computed in the same vein. Hence, 
\begin{align}
	\label{appendix integral I 2} I &= \int d^2x\: e^{i x \cdot \Delta} \int \frac{dq_1}{2\pi} \frac{i e^{-ix^1 q_1}}{2 \omega(q)} \left\{- \frac{i}{2} \left[ \left( \dslash{q} + m \right) e^{-i x^0 q_0}\Big|_{q_0 = \omega(q)} - \left( \dslash{q} + m \right) e^{-i x^0 q_0}\Big|_{q_0 = -\omega(q)}\right] \right. + \nonumber \\
	&+ \left. \fint \frac{dq_0}{2\pi}\: \left( \dslash{q} + m \right) e^{-i x^0 q_0} \left[ \frac{1}{q_0 - \omega(q)} - \frac{1}{q_0 + \omega(q)} \right] \right\} 4 \pi^2 \delta^{(2)}(\mathbf{k} - \mathbf{p}). 
\end{align}

As stated above, the terms in the first line of (\ref{appendix integral I 2}) will contribute to the delta terms in the decomposition (\ref{delta + pv split}). In order to conclude this, it is only necessary to group the exponentials together, and realize that one can safely exchange the order of integrations to obtain a delta function from the integration over $x$. Namely, 
\begin{align}
	\label{appendix integral I delta} \int d^2x\: e^{i x \cdot \Delta} \int \frac{dq_1}{2\pi} \frac{i e^{-ix^1 q_1}}{2 \omega(q)} \left\{- \frac{i}{2} \left[ \left( \dslash{q} + m \right) e^{-i x^0 q_0}\Big|_{q_0 = \omega(q)} - \left( \dslash{q} + m \right) e^{-i x^0 q_0}\Big|_{q_0 = -\omega(q)}\right] \right\} = \nonumber \\
	= \frac{\dslash{\Delta} + m}{4 \omega (\Delta)}\: 2 \pi \left[ \delta \Big( \Delta_0 - \omega(\Delta) \Big) - \delta \Big( \Delta_0 + \omega(\Delta) \Big) \right]. 
\end{align}

Deriving the \rm{p.v.} contribution demands more work. First, let us denote the principal value integral from (\ref{appendix integral I 2}) simply as $ \dslash{I}$ and note that: 
\begin{align}
	\left(\dslash{q} + m\right) \left[ \frac{1}{q_0 - \omega(q)} - \frac{1}{q_0 + \omega(q)} \right] &= \gamma^0 \omega(q) \left[ \frac{1}{q_0 - \omega(q)} + \frac{1}{q_0 + \omega(q)} \right] + \nonumber \\
	&+ \left( q_1\gamma^1 +m \right) \left[ \frac{1}{q_0 - \omega(q)} - \frac{1}{q_0 + \omega(q)} \right]. 
\end{align}
Hence, if we use the identities: 
\begin{align}
	\fint \frac{dq_0}{2 \pi} \: e^{-ix^0 q_0} \left[ \frac{1}{q_0 - \omega(q)} - \frac{1}{q_0 + \omega(q)} \right] &= - \frac{2}{\pi} \sin \Big( \omega(q) x^0 \Big) \int_{\epsilon}^{\infty} \frac{dz}{z} \sin \Big( x^0 z \Big), \\
	\fint \frac{dq_0}{2 \pi} \: e^{-ix^0 q_0} \left[ \frac{1}{q_0 - \omega(q)} + \frac{1}{q_0 + \omega(q)} \right] &= - \frac{2i}{\pi} \cos \Big( \omega(q) x^0 \Big) \int_{\epsilon}^{\infty} \frac{dz}{z} \sin \Big( x^0 z \Big), 
\end{align}
where we left implicit the $\epsilon \to 0$ limit, we can write: 
\begin{align}
	\dslash{I} &= \int d^2x \: e^{i x \cdot \Delta} \int \frac{dq_1}{2 \pi} \: \frac{i e^{-ix^1 q_1}}{2 \omega(q)} \left\{ -\frac{2i}{\pi} \gamma^0 \omega(q) \cos \Big( \omega(q) x^0 \Big) - \frac{2}{\pi} \left(q_1 \gamma^1 + m \right) \sin \Big( \omega(q) x^0 \Big) \right\} \cdot \nonumber \\
	&\cdot \int_{\epsilon}^{\infty} \frac{dz}{z} \sin \Big( x^0 z \Big) \: 4 \pi \delta^{(2)} \left( \mathbf{k} - \mathbf{p} \right) \nonumber \\
	&= \frac{i}{8 \pi^2} \int d^2x \int dq_1 \frac{e^{ix^1 \left( \Delta_1 - q_1 \right)}}{\omega(q)} \left[ \int_{-\infty}^{-\epsilon} + \int^{\infty}_{\epsilon} \right] \frac{dz}{z}\: \bigg\{ \Big[ \omega(q) \gamma^0 - q_1\gamma^1 - m\Big] e^{ix^0 \left[ \Delta_0 + \omega(q) -z \right]} \bigg. + \nonumber \\
	&+ \bigg. \Big[ \omega(q) \gamma^0 + q_1\gamma^1 +m\Big] e^{ix^0 \left[ \Delta_0 - \omega(q) -z \right]} \bigg\} \: 4 \pi^2 \delta^{(2)} \left( \mathbf{k} - \mathbf{p} \right). 
\end{align}

Next, we exchange the order of integrations so as to evaluate the integrals over $x^1$ and $x^0$ first. The integration over $x^1$ factorizes and clearly yields a delta function, which can then be used to integrate over $q_1$, casting $q_1 = \Delta_1$. The situation involving the integration over $x^0$ is, however, more delicate. Here the limit $\epsilon \to 0$ plays a paramount role, as it removes the point $z = 0$ from the integration domain, thus, removing the $\sim \frac{1}{z}$ singularity, and we can safely proceed as before. Namely, integrate over $x^0$ to obtain a delta function, which in turn allows us to perform the integration over $z$. Concluding, thus, that: 
\begin{align}
	\label{appendix integral I pv} \dslash{I} &= \frac{i}{2 \omega (\Delta)} \left\{ \frac{\omega(\Delta) \gamma^0 - \Delta_1 \gamma^1 - m}{\Delta_0 + \omega(\Delta)} + \frac{\omega(\Delta) \gamma^0 + \Delta_1 \gamma^1 + m}{\Delta_0 - \omega(\Delta)} \right\} \: 4 \pi^2 \delta^{(2)} \left( \mathbf{k} - \mathbf{p} \right) \nonumber \\
	&= \frac{i \left( \dslash{\Delta} +m \right)}{\Delta^2 - m^2} \: 4 \pi^2 \delta^{(2)} \left( \mathbf{k} - \mathbf{p} \right). 
\end{align}
Finally, by adding (\ref{appendix integral I delta}) and (\ref{appendix integral I pv}) together, we obtain the desired result, (\ref{3p I}). 

We stress, however, that a careful analysis of whether the factor $\Delta_0 \pm \omega(\Delta)$ vanishes is of great importance, as it may imply divergent or discontinuous scattering amplitudes (see section \ref{continuity} for the discussion of this subtlety). The plus case is easier to understand, since we deal only with pseudo-particles with positive energy, and thus, if we impose the mass-shell condition\footnote{Remember that $\Delta = p^1 + p^2 - k^1$.}, one easily sees that such a polynomial has no real roots. This is obviously not the case for the on-shell polynomial coming from $\Delta_0 - \omega(\Delta)$, which can be solved, say for $k^1$, giving $k^1 = p^1$ or $k^1 = p^2$. Although we can still perform the integrations over $x^0$ and $z$ as we did above, even if $\Delta_0 - \omega(\Delta) = 0$, the result is obviously not the same as (\ref{appendix integral I pv}). In fact, it changes in such a way as to avoid any divergencies coming from the vanishing of the denominator. There remains only the question about the continuity of the forthcoming scattering amplitudes as we approach the integrability point. But as discussed in the main text, this issue can be dismissed by considering localized wave-packet distributions, because in this case the condition $\Delta_0 - \omega(\Delta) = 0$ is never satisfied.

\section{Tree-level factorizability computational details}\label{Factorizable Tree-Level Details}

In this appendix we collect all the additional formulae needed for computing the cancellation of the spurious contributions that prevented $S$-matrix factorization. The factors $A_i \left( \boldsymbol{\eta}, \boldsymbol{\theta} \right)$ which provide the decomposition (\ref{G decomposition}) of $G \left( \boldsymbol{\eta}, \boldsymbol{\theta} \right)$ are: 
 \begin{align}
    A_1 \left( \boldsymbol{\eta}, \boldsymbol{\theta} \right) &= \cosh
    \left( \frac{\eta_2 - \eta_3}{2} \right) \cosh \left(
      \frac{\theta_1 - \theta_2}{2} \right) \cosh \left(
      \frac{\theta_2 - \theta_3}{2} \right) \csch \left( \frac{\eta_1
        - \theta_1}{2} \right) \sinh \left( \frac{\eta_2 - \eta_3}{2}
    \right) \cdot \nonumber \\ \displaybreak[0] 
    &\cdot \sinh \left( \frac{2\eta_1 - \theta_1 - \theta_2}{2}
    \right) \sinh \left( \frac{\eta_2 + \eta_3 - 2\theta_3}{2}
    \right), \label{1-loop pv A1}\\ 
    A_2 \left( \boldsymbol{\eta}, \boldsymbol{\theta} \right) &= -
    \cosh \left( \frac{\eta_2 - \eta_3}{2} \right) \cosh \left(
      \frac{\theta_1 - \theta_2}{2} \right) \cosh \left(
      \frac{\theta_2 - \theta_3}{2} \right) \csch \left( \frac{\eta_1
        - \theta_1}{2} \right) \cdot \nonumber \\ \displaybreak[0] 
    &\cdot \sinh^2 \left( \frac{\eta_2 - \eta_3}{2} \right) \sinh
    \left( \frac{2\eta_1 - \theta_1 - \theta_2}{2}
    \right), \label{1-loop pv A2}\\ 
    A_3 \left( \boldsymbol{\eta}, \boldsymbol{\theta} \right) &= \cosh
    \left( \frac{\eta_2 - \eta_3}{2} \right) \cosh \left(
      \frac{\theta_1 - \theta_2}{2} \right) \cosh \left(
      \frac{\theta_2 - \theta_3}{2} \right) \csch \left( \frac{\eta_1
        - \theta_1}{2} \right) \sinh \left( \frac{\eta_2 - \eta_3}{2}
    \right) \cdot \nonumber \\ \displaybreak[0] 
    &\cdot \sinh \left( \frac{\theta_1 - \theta_2}{2} \right) \sinh
    \left( \frac{\eta_2 + \eta_3 - 2 \theta_3}{2}
    \right), \label{1-loop pv A3}\\ 
    A_4 \left( \boldsymbol{\eta}, \boldsymbol{\theta} \right) &= -
    \cosh \left( \frac{\eta_2 - \eta_3}{2} \right) \cosh \left(
      \frac{\theta_1 - \theta_2}{2} \right) \cosh \left(
      \frac{\theta_2 - \theta_3}{2} \right) \csch \left( \frac{\eta_1
        - \theta_1}{2} \right) \cdot \nonumber \\ \displaybreak[0] 
    &\cdot \sinh^2 \left( \frac{\eta_2 - \eta_3}{2} \right) \sinh
    \left( \frac{\theta_1 - \theta_2}{2} \right), \label{1-loop pv
      A4}\\ 
    A_5 \left( \boldsymbol{\eta}, \boldsymbol{\theta} \right) &= -
    \cosh^2 \left( \frac{\eta_2 - \eta_3}{2} \right) \cosh \left(
      \frac{\theta_1 - \theta_2}{2} \right) \csch \left( \frac{\eta_1
        - \theta_1}{2} \right) \sinh \left( \frac{\theta_2 -
        \theta_3}{2} \right) \cdot \nonumber \\ \displaybreak[0] 
    &\cdot \sinh \left( \frac{2\eta_1 - \theta_1 - \theta_2}{2}
    \right) \sinh \left( \frac{\eta_2 + \eta_3 - 2\theta_3}{2}
    \right), \label{1-loop pv A5}\\ 
    A_6 \left( \boldsymbol{\eta}, \boldsymbol{\theta} \right) &=
    \cosh^2 \left( \frac{\eta_2 - \eta_3}{2} \right) \cosh \left(
      \frac{\theta_1 - \theta_2}{2} \right) \csch \left( \frac{\eta_1
        - \theta_1}{2} \right) \sinh \left( \frac{\eta_2 - \eta_3}{2}
    \right) \cdot \nonumber \\ \displaybreak[0] 
    &\cdot \sinh \left( \frac{\theta_2 - \theta_3}{2} \right) \sinh
    \left( \frac{2\eta_1 - \theta_1 - \theta_2}{2}
    \right), \label{1-loop pv A6}\\ 
    A_7 \left( \boldsymbol{\eta}, \boldsymbol{\theta} \right) &=-
    \cosh^2 \left( \frac{\eta_2 - \eta_3}{2} \right) \cosh \left(
      \frac{\theta_1 - \theta_2}{2} \right) \csch \left( \frac{\eta_1
        - \theta_1}{2} \right) \sinh \left( \frac{\theta_1 -
        \theta_2}{2} \right) \cdot \nonumber \\ \displaybreak[0]  
    &\cdot \sinh \left( \frac{\theta_2 - \theta_3}{2} \right) \sinh
    \left( \frac{\eta_2 + \eta_3 - 2\theta_3}{2}
    \right), \label{1-loop pv A7}\\ 
    A_8 \left( \boldsymbol{\eta}, \boldsymbol{\theta} \right) &=
    \cosh^2 \left( \frac{\eta_2 - \eta_3}{2} \right) \cosh \left(
      \frac{\theta_1 - \theta_2}{2} \right) \csch \left( \frac{\eta_1
        - \theta_1}{2} \right) \sinh \left( \frac{\eta_2 - \eta_3}{2}
    \right) \cdot \nonumber \\ \displaybreak[0] 
    &\cdot \sinh \left( \frac{\theta_1 - \theta_2}{2} \right) \sinh
    \left( \frac{ \theta_2 - \theta_3}{2} \right).\label{1-loop pv A8}  
  \end{align} 

Bellow we give the explicit expressions for the non-vanishing action of the antisymmetrizer on the factors $A_i \left( \boldsymbol{\eta}, \boldsymbol{\theta} \right)$: 
\begin{align}
    \left( 3! \right)^2 \mathcal{A}_{\theta, \eta} \left[ A_1 \left(
        \boldsymbol{\eta}, \boldsymbol{\theta} \right) \right] &=
    \frac{1}{4} \bigg\{ \csch \left[ \frac{\eta_1 - \theta_1}{2}
    \right] \csch \left[ \frac{\eta_1 - \theta_2}{2} \right] \csch
    \left[ \frac{\eta_1 - \theta_3}{2} \right] \sinh \left( \eta_2 -
      \eta_3 \right) \cdot \nonumber \\ 
    &\cdot \bigg[ \bigg. 3 \sinh \left( \frac{2 \eta_1 - \eta_2 -
        \eta_3}{2} \right) - \sinh \left( \frac{4 \eta_1 +\eta_2 +
        \eta_3 - 2 \theta_1 - 2 \theta_2 -2\theta_3}{2} \right) -
    \nonumber \\ 
    &- \sinh \left( \frac{4 \eta_1 - \eta_2 - \eta_3 - 2 \theta_1}{2} \right) - \sinh \left( \frac{4 \eta_1 - \eta_2 - \eta_3 - 2 \theta_2}{2} \right) - \nonumber \\
    &- \sinh \left( \frac{4 \eta_1 - \eta_2 - \eta_3 - 2 \theta_3}{2}
    \right) + \sinh \left( \frac{2 \eta_1 + \eta_2 + \eta_3 - 2
        \theta_1 - 2 \theta_2}{2} \right) + \nonumber \\ 
    &+ \sinh \left( \frac{2 \eta_1 + \eta_2 + \eta_3 - 2 \theta_1 - 2
        \theta_3}{2} \right) + \sinh \left( \frac{2 \eta_1 + \eta_2 +
        \eta_3 - 2 \theta_2 - 2 \theta_3}{2} \right) + \nonumber \\ 
    &+ \sinh \left( \frac{2 \eta_1 - \eta_2 - \eta_3 + 2\theta_1 -
        2\theta_2}{2} \right) + \sinh \left( \frac{2 \eta_1 - \eta_2 -
        \eta_3 - 2\theta_1 + 2\theta_2}{2} \right) + \nonumber \\ 
    &+ \sinh \left( \frac{2 \eta_1 - \eta_2 - \eta_3 + 2\theta_1 -
        2\theta_3}{2} \right) + \sinh \left( \frac{2 \eta_1 - \eta_2 -
        \eta_3 -2 \theta_1 + 2 \theta_3}{2} \right) + \nonumber \\ 
    &+ \sinh \left( \frac{2 \eta_1 - \eta_2 - \eta_3 + 2\theta_2 -
        2\theta_3}{2} \right) + \sinh \left( \frac{2 \eta_1 - \eta_2 -
        \eta_3 - 2\theta_2 + 2 \theta_3}{2} \right) \bigg. \bigg] +
    \nonumber \displaybreak[0] \\ 
    &+ \csch \left[ \frac{\eta_2 - \theta_1}{2} \right] \csch \left[
      \frac{\eta_2 - \theta_2}{2} \right] \csch \left[ \frac{\eta_2 -
        \theta_3}{2} \right] \sinh \left( \eta_1 - \eta_3 \right)
    \cdot \nonumber \\ 
    &\cdot \bigg[ \bigg. 3 \sinh \left( \frac{ \eta_1 - 2 \eta_2 +
        \eta_3}{2} \right) + \sinh \left( \frac{\eta_1 + 4\eta_2 +
        \eta_3 - 2 \theta_1 - 2 \theta_2 -2\theta_3}{2} \right) -
    \nonumber \\ 
    &- \sinh \left( \frac{ \eta_1 - 4\eta_2 + \eta_3 - 2 \theta_1}{2}
    \right) - \sinh \left( \frac{ \eta_1 - 4\eta_2 + \eta_3 - 2
        \theta_2}{2} \right) - \nonumber \\ 
    &- \sinh \left( \frac{ \eta_1 - 4\eta_2 + \eta_3 - 2 \theta_3}{2}
    \right) - \sinh \left( \frac{ \eta_1 + 2\eta_2 + \eta_3 - 2
        \theta_1 - 2 \theta_2}{2} \right) - \nonumber \\ 
    &- \sinh \left( \frac{ \eta_1 + 2\eta_2 + \eta_3 - 2 \theta_1 - 2
        \theta_3}{2} \right) - \sinh \left( \frac{ \eta_1 + 2\eta_2 +
        \eta_3 - 2 \theta_2 - 2 \theta_3}{2} \right) + \nonumber \\ 
    &+ \sinh \left( \frac{ \eta_1 - 2 \eta_2 + \eta_3 + 2\theta_1 -
        2\theta_2}{2} \right) + \sinh \left( \frac{ \eta_1 - 2 \eta_2
        + \eta_3 - 2\theta_1 + 2 \theta_2}{2} \right) + \nonumber \\ 
    &+ \sinh \left( \frac{ \eta_1 - 2 \eta_2 + \eta_3 + 2\theta_1 -
        2\theta_3}{2} \right) + \sinh \left( \frac{ \eta_1 - 2 \eta_2
        + \eta_3 - 2\theta_1 + 2\theta_3}{2} \right) + \nonumber \\ 
    \displaybreak[0]
    &+ \sinh \left( \frac{ \eta_1 - 2\eta_2 + \eta_3 + 2\theta_2 -
        2\theta_3}{2} \right) + \sinh \left( \frac{ \eta_1 - 2\eta_2 +
        \eta_3 - 2\theta_2 + 2\theta_3}{2} \right) \bigg. \bigg] -
    \nonumber \\ 
    &- \csch \left[ \frac{\eta_3 - \theta_1}{2} \right] \csch \left[
      \frac{\eta_3 - \theta_2}{2} \right] \csch \left[ \frac{\eta_3 -
        \theta_3}{2} \right] \sinh \left( \eta_1 - \eta_2 \right)
    \cdot \nonumber \\ 
    &\cdot \bigg[ \bigg. 3 \sinh \left( \frac{ \eta_1 + \eta_2 -2
        \eta_3}{2} \right) + \sinh \left( \frac{ \eta_1 +\eta_2 +
        4\eta_3 - 2 \theta_1 - 2 \theta_2 -2\theta_3}{2} \right) -
    \nonumber \\ 
    &- \sinh \left( \frac{ \eta_1 + \eta_2 - 4 \eta_3 + 2 \theta_1}{2}
    \right) - \sinh \left( \frac{ \eta_1 + \eta_2 - 4 \eta_3 + 2
        \theta_2}{2} \right) - \nonumber \\ 
    &- \sinh \left( \frac{4 \eta_1 + \eta_2 - 4 \eta_3 + 2
        \theta_3}{2} \right) - \sinh \left( \frac{ \eta_1 + \eta_2 +
        2\eta_3 - 2 \theta_1 - 2 \theta_2}{2} \right) - \nonumber \\ 
    &- \sinh \left( \frac{ \eta_1 + \eta_2 + 2\eta_3 - 2 \theta_1 - 2
        \theta_3}{2} \right) - \sinh \left( \frac{ \eta_1 + \eta_2 +
        2\eta_3 - 2 \theta_2 - 2 \theta_3}{2} \right) + \nonumber \\ 
    &+ \sinh \left( \frac{ \eta_1 + \eta_2 - 2 \eta_3 + 2\theta_1 -
        2\theta_2}{2} \right) + \sinh \left( \frac{ \eta_1 + \eta_2 -
        2 \eta_3 - 2\theta_1 + 2\theta_2}{2} \right) + \nonumber \\ 
    &+ \sinh \left( \frac{ \eta_1 + \eta_2 - 2 \eta_3 + 2\theta_1 -
        2\theta_3}{2} \right) + \sinh \left( \frac{ \eta_1 + \eta_2 -
        2 \eta_3 - 2\theta_1 + 2\theta_3}{2} \right) + \nonumber \\ 
    &+ \sinh \left( \frac{ \eta_1 + \eta_2 - 2 \eta_3 + 2\theta_2 -
        2\theta_3}{2} \right) + \sinh \left( \frac{ \eta_1 + \eta_2 -
        2 \eta_3 - 2\theta_2 + 2\theta_3}{2} \right) \bigg. \bigg]
    \bigg. \bigg\} \cdot \nonumber \\ 
    &\cdot \sinh \left( \frac{\theta_1 - \theta_2}{2} \right) \sinh
    \left( \frac{\theta_1 - \theta_3}{2} \right) \sinh \left(
      \frac{\theta_2 - \theta_3}{2} \right), \displaybreak[1] \\ 
    \left( 3! \right)^2 \mathcal{A}_{\theta, \eta} \left[ A_3 \left(
        \boldsymbol{\eta}, \boldsymbol{\theta} \right) \right] &=
    \frac{1}{4} \bigg\{ - \csch \left( \frac{\eta_3 - \theta_1}{2}
    \right) \csch \left( \frac{\eta_3 - \theta_2}{2} \right) \csch
    \left( \frac{\eta_3 - \theta_3}{2} \right) \sinh \left( \eta_1 -
      \eta_2 \right) \bigg. \cdot \nonumber \\ 
    &\cdot \bigg[ \sinh \left( \frac{\eta_1 + \eta_2 - 2
        \eta_3}{2} \right) + \sinh \left( \frac{\eta_1 + \eta_2 -
        2 \theta_1}{2} \right) + \sinh \left( \frac{\eta_1 +
        \eta_2 - 2 \theta_2}{2} \right) + \bigg. \nonumber \\ 
    & + \sinh \left( \frac{\eta_1 + \eta_2 - 2 \theta_3}{2}
    \right) - \sinh \left( \frac{\eta_1 + \eta_2 + 2 \eta_3 -2
        \theta_1 - 2\theta_2}{2} \right) - \nonumber \\ 
    &- \sinh \left( \frac{\eta_1 + \eta_2 + 2 \eta_3 -2 \theta_1 -
        2\theta_3}{2} \right) - \sinh \left( \frac{\eta_1 + \eta_2
        + 2 \eta_3 -2 \theta_2 - 2\theta_3}{2} \right) \nonumber +
    \\ 
    &+ \sinh \left( \frac{\eta_1 + \eta_2 - 2 \theta_1 -2 \theta_2
        + 2\theta_3}{2} \right) + \sinh \left( \frac{\eta_1 +
        \eta_2 - 2 \theta_1 +2 \theta_2 - 2\theta_3}{2} \right) +
    \nonumber \\ 
    &+ \sinh \left( \frac{\eta_1 + \eta_2 + 2 \theta_1 -2 \theta_2
        - 2\theta_3}{2} \right) \bigg. \bigg] + \nonumber
    \displaybreak[0] \\  
    &+ \csch \left( \frac{\eta_2 - \theta_1}{2} \right) \csch
    \left( \frac{\eta_2 - \theta_2}{2} \right) \csch \left(
      \frac{\eta_2 - \theta_3}{2} \right) \sinh \left( \eta_1 -
      \eta_3 \right) \bigg. \cdot \nonumber \\ 
    &\cdot \bigg[ \sinh \left( \frac{\eta_1 - 2 \eta_2 +
        \eta_3}{2} \right) + \sinh \left( \frac{\eta_1 + \eta_3 -
        2 \theta_1}{2} \right) + \sinh \left( \frac{\eta_1 +
        \eta_3 - 2 \theta_2}{2} \right) + \bigg. \nonumber \\ 
    & + \sinh \left( \frac{\eta_1 + \eta_3 - 2 \theta_3}{2}
    \right) - \sinh \left( \frac{\eta_1 + 2\eta_2 + \eta_3 -2
        \theta_1 - 2\theta_2}{2} \right) - \nonumber \\ 
    &- \sinh \left( \frac{\eta_1 + 2\eta_2 + \eta_3 -2 \theta_1 -
        2\theta_3}{2} \right) - \sinh \left( \frac{\eta_1 +
        2\eta_2 + \eta_3 -2 \theta_2 - 2\theta_3}{2} \right)
    \nonumber + \\ 
    &+ \sinh \left( \frac{\eta_1 + \eta_3 - 2 \theta_1 -2 \theta_2
        + 2\theta_3}{2} \right) + \sinh \left( \frac{\eta_1 +
        \eta_3 - 2 \theta_1 +2 \theta_2 - 2\theta_3}{2} \right) +
    \nonumber \\ 
    &+ \sinh \left( \frac{\eta_1 + \eta_3 + 2 \theta_1 -2 \theta_2
        - 2\theta_3}{2} \right) \bigg. \bigg] + \nonumber \\ 
    &+ \csch \left( \frac{\eta_1 - \theta_1}{2} \right) \csch
    \left( \frac{\eta_1 - \theta_2}{2} \right) \csch \left(
      \frac{\eta_1 - \theta_3}{2} \right) \sinh \left( \eta_2 -
      \eta_3 \right) \bigg. \cdot \nonumber \\ 
    &\cdot \bigg[ \sinh \left( \frac{2 \eta_1 - \eta_2 -
        \eta_3}{2} \right) - \sinh \left( \frac{\eta_2 + \eta_3 -
        2 \theta_1}{2} \right) - \sinh \left( \frac{\eta_2 +
        \eta_3 - 2 \theta_2}{2} \right) - \bigg. \nonumber \\ 
    & - \sinh \left( \frac{\eta_2 + \eta_3 - 2 \theta_3}{2}
    \right) + \sinh \left( \frac{2 \eta_1 + \eta_2 + \eta_3 -2
        \theta_1 - 2\theta_2}{2} \right) + \nonumber \\ 
    &+ \sinh \left( \frac{2\eta_1 + \eta_2 + \eta_3 -2 \theta_1 -
        2\theta_3}{2} \right) + \sinh \left( \frac{2 \eta_1 +
        \eta_2 + \eta_3 -2 \theta_2 - 2\theta_3}{2} \right)
    \nonumber - \\ 
    &- \sinh \left( \frac{\eta_2 + \eta_3 - 2 \theta_1 -2 \theta_2
        + 2\theta_3}{2} \right) - \sinh \left( \frac{\eta_2 +
        \eta_3 - 2 \theta_1 +2 \theta_2 - 2\theta_3}{2} \right) -
    \nonumber \\ 
    &- \sinh \left( \frac{\eta_2 + \eta_3 + 2 \theta_1 -2 \theta_2
        - 2\theta_3}{2} \right) \bigg. \bigg] \bigg. \bigg\} \cdot
    \nonumber \\ 
    &\cdot \sinh \left( \frac{\theta_1 - \theta_2}{2} \right)
    \sinh \left( \frac{\theta_1 - \theta_3}{2} \right) \sinh
    \left( \frac{\theta_2 - \theta_3}{2} \right), \displaybreak[1]\\ 
    \left( 3! \right)^2 \mathcal{A}_{\theta, \eta} \left[ A_6
      \left( \boldsymbol{\eta}, \boldsymbol{\theta} \right)
    \right] &= \frac{1}{4} \bigg\{ \csch \left( \frac{\eta_2 -
        \eta_3}{2} \right) \csch \left( \frac{\eta_1 -
        \theta_1}{2} \right) \csch \left( \frac{\eta_1 -
        \theta_2}{2} \right) \csch \left( \frac{\eta_1 -
        \theta_3}{2} \right) \cdot \bigg. \nonumber \\ 
    &\cdot \sinh^2 \left( \eta_2 - \eta_3 \right) \Big[ \sinh
    \left( \eta_1 - \eta_2 \right) + \sinh \left( \eta_1 - \eta_3
    \right) \Big] - \nonumber \displaybreak[0] \\ 
    &- \csch \left( \frac{\eta_1 - \eta_3}{2} \right) \csch \left(
      \frac{\eta_2 - \theta_1}{2} \right) \csch \left(
      \frac{\eta_2 - \theta_2}{2} \right) \csch \left(
      \frac{\eta_2 - \theta_3}{2} \right) \cdot \nonumber \\ 
    &\cdot \sinh^2 \left( \eta_1 - \eta_3 \right) \Big[ \sinh
    \left( \eta_2 - \eta_1 \right) + \sinh \left( \eta_2 - \eta_3
    \right) \Big] + \nonumber \displaybreak[0] \\ 
    &+ \csch \left( \frac{\eta_1 - \eta_2}{2} \right) \csch \left(
      \frac{\eta_3 - \theta_1}{2} \right) \csch \left(
      \frac{\eta_3 - \theta_2}{2} \right) \csch \left(
      \frac{\eta_3 - \theta_3}{2} \right) \cdot \nonumber \\ 
    &\cdot \bigg. \sinh^2 \left( \eta_1 - \eta_2 \right) \Big[
    \sinh \left( \eta_3 - \eta_1 \right) + \sinh \left( \eta_3 -
      \eta_2 \right) \Big] \bigg\} \cdot \nonumber \\ 
    &\cdot \sinh \left( \frac{\theta_1 - \theta_2}{2} \right)
    \sinh \left( \frac{\theta_1 - \theta_3}{2} \right) \sinh
    \left( \frac{\theta_2 - \theta_3}{2} \right).  
  \end{align}

\clearpage 
\bibliographystyle{JHEP3} 
\bibliography{AAF_final}

\end{document}